\documentclass[12pt]{article}
\usepackage{graphicx,psfrag,epsf}
\usepackage{enumerate}
\usepackage{natbib}
\usepackage{url} 
\usepackage{hyperref}
\usepackage{amsmath,amssymb,amsfonts,bm,bbm,amsthm} 
\usepackage{color}
\usepackage{enumitem}
\usepackage{mathtools}
\usepackage{algpseudocode}
\usepackage{algorithmicx, algorithm}
\usepackage{booktabs}
\usepackage{subcaption}
\usepackage{xr}
\usepackage[T1]{fontenc}
\usepackage[utf8]{inputenc}

\numberwithin{equation}{section}
\newtheorem{theorem}{Theorem}[section]

\newtheorem{corollary}[theorem]{Corollary}

\newtheorem{proposition}[theorem]{Proposition}

\newtheorem{remark}[theorem]{Remark}
\newtheorem{example}[theorem]{Example}

\def\bsX{\boldsymbol{X}}
\def\bsx{\boldsymbol{x}}

\def\bsz{\boldsymbol{z}}
\def\bsbeta{\boldsymbol\beta}

\def\bsSigma{\boldsymbol\Sigma}

\def\bsI{\boldsymbol{I}}
\def\bsb{\boldsymbol{b}}
\def\bsy{\boldsymbol{y}}

\def\R{\mathbb R}
\def\E{\mathbb E}
\def\P{\mathbb P}
\def\argmin{\operatornamewithlimits{argmin}}

\def\1{\mathbbm 1}

\newcommand{\blind}{1}

\begin{document}

\def\spacingset#1{\renewcommand{\baselinestretch}%
{#1}\small\normalsize} \spacingset{1}


\if1\blind
{
  \title{\bf Selection and Estimation Optimality in High Dimensions with the TWIN Penalty}
  \author{Xiaowu Dai \thanks{Equal contribution.}  \thanks{Xiaowu Dai is a PhD student at the Department of Statistics, University of Wisconsin-Madison (E-mail: \href{mailto:xdai26@wisc.edu}{xdai26@wisc.edu}). Supported in part by NSF Grant DMS-1308877.} \ \ 
    and \ 
    Jared Huling \footnotemark[1]  \thanks{Jared Huling is an Assistant Professor at the Department of Statistics, The Ohio State University (E-mail: \href{mailto:huling.7@osu.edu}{huling.7@osu.edu}).}}
  \date{}
  \maketitle
} \fi

\if0\blind
{
  \bigskip
  \bigskip
  \bigskip
  \begin{center}
    {\Large \bf The View from The Two Mountains:  \\ Controlling False Discoveries in High-dimensional Regression}
\end{center}
  \medskip
} \fi


\begin{abstract}
	We introduce a novel class of variable selection penalties called TWIN, which provides sensible data-adaptive penalization. Under a linear sparsity regime and random Gaussian designs we show that penalties in the TWIN class have a high probability of selecting the correct model and furthermore result in minimax optimal estimators. The general shape of penalty functions in the TWIN class is the key ingredient to its desirable properties and results in improved theoretical and empirical performance over existing penalties. In this work we introduce two examples of TWIN penalties that admit simple and efficient coordinate descent algorithms, making TWIN practical in large data settings. We demonstrate in challenging and realistic simulation settings with high correlations between active and inactive variables that TWIN has high power in variable selection while controlling the number of false discoveries, outperforming standard penalties.
\end{abstract}

\noindent%
{ Keywords: Sparse regression; Penalized likelihood; Variable selection; False discovery rate; Minimax optimality.}  
\vfill

\newpage
\spacingset{1.45} 

\section{Introduction}
\label{sec:intro}

Discovering relevant relationships between a large number of variables and an outcome continues to be an eminently challenging problem in statistics and a major interest in a wide variety of scientific disciplines.   Decades of research has focused on variable selection techniques to identify relevant variables.
Among these techniques, penalized regression-based methods such as the Lasso \citep{tibshirani1996}, smoothly-clipped absolute deviation (SCAD) \citep{fan2001}, and the minimax concave penalty (MCP) \citep{zhang2010} have been widely explored, as they often perform well in practice, have computational advantages, and possess desirable variable selection properties. However, selection consistency results for penalized methods often require the imposition of relatively extreme levels of sparsity on the data generating mechanism and thus may not accurately describe real world data. 
For example, when modeling health outcomes of patients, such as hospitalization risk or human phenotypes, the relevant risk factors may be highly varied and numerous. As human biology is extraordinarily complex, it is sensible that more relevant predictors may be included when an increasing amount of genetic or microbiome information is leveraged, especially when considering gene-gene, gene-environment, or microbiome-environment interactions \citep{nadeau2006genetics, martin2007genetic, bull2014part, shreiner2015gut}. As such, methodological and theoretical advances in variable selection commensurate with this possibility are needed.

In this paper we seek to address this gap with a novel class of penalties. The proposed penalty  class results in estimators that are provably selection consistent and asymptotically minimax in high-dimensional scenarios under linear sparsity and relatively weak assumptions regarding the data-generating mechanism. We call our penalty class the two mountains penalty class, or TWIN ({\bf TW}o mounta{\bf IN}s) for short, as the shape of the penalty function resembles two mountains centered around the origin.  The general shape of the two mountains penalty class makes it amenable to controlling the false discovery rate of variable selections (FDR) while retaining high power of selection and is thus instrumental to its desirable selection properties. Furthermore, the shape of TWIN penalty functions, illustrated in Figure \ref{fig:two_mountains_example}, results in sensible data-adaptive penalization where larger coefficients are subjected to attenuated penalization. Throughout this paper we show that this general pattern of penalization yields advantageous selection and estimation properties. 
Extensive simulations buttress our theoretical results and demonstrate the superior finite sample selection and estimation properties of our penalty in scenarios with strong correlations between relevant and irrelevant variables.

The core of this paper centers around the ubiquitous linear model, which posits that the relationship between a set of predictors and a response variable has the following linear form:
\begin{equation}\label{eqn:linear_model}
\bsy = \bsX\bsbeta + \bsz,
\end{equation}
where $\bsy \in \R^n$ is a vector of responses, $\bsX \equiv (\bsx_1,\ldots,\bsx_p) \in \R^{n \times p}$ is a random matrix with each column representing samples of a particular predictor, $\bsbeta = (\beta_1, \dots, \beta_p)' \in \R^p$ is a vector which relates the predictors to a mean response value, and $\bsz\sim N(0, \sigma^2\bsI_n)$ is an error term independent of $\bsX$. 
We adopt the familiar penalized regression framework, wherein sparse estimates $\widehat{\bsbeta}$ of $\bsbeta$ are achieved by minimizing a penalized least squares objective with penalty $P(\cdot)$:
\begin{equation}
\label{eqn:penalized_least_squares}
\widehat{\bsbeta} =  \underset{\bsb\in\R^p}{\arg\min}\left\{\frac{1}{2}||\bsy - \bsX\bsb||^2 +\sum_{j=1}^pP(|b_j|)\right\}.
\end{equation}
The Lasso falls under this framework with $P(|b|) = |b|$. The focus of this paper is on a new class of penalty functions $P(\cdot)$, which will be introduced in Section \ref{sec:metho}. 

We highlight three main contributions of this work:
\begin{enumerate}
\item We propose a novel class of penalty functions for variable selection, which provide data-adaptive penalization in a manner which results empirically in favorable selection and prediction performance. We provide two examples of the penalty class which are amenable to computationally efficient algorithms.
\item We provide selection consistency results for the proposed class of penalty functions in both the high dimensional ($p > n$) and low-dimensional settings under linear sparsity. Similar to SLOPE \citep{bogdan2015}, our penalty admits a finite sample bound for the FDR under orthogonality and is thus a candidate for future study of FDR control under more general designs. 
\item We establish new minimax optimal risk under the linear sparsity. Moreover, we show that TWIN estimators are minimax optimal for both orthogonal and random designs.
\end{enumerate}

The remainder of this paper is organized as follows. 
 We introduce our proposed class of penalty functions in Section \ref{sec:metho}. In Section \ref{sec:selectionconsis} we study the key selection properties of the TWIN penalty and in Section \ref{sec:estimationproperties} we present minimax optimality results. Section \ref{sec:simulation} investigates the numerical properties of the TWIN penalty in comparison with other standard penalties using extensive simulation studies. In Section \ref{sec:application} we analyze a microarray study relating gene expression levels to a phenotype in mice  with the TWIN penalty. A summary and some discussion are given in Section \ref{sec:discussion}.

\section{Methodology}
\label{sec:metho}

\subsection{The TWIN penalty class and examples}\label{sec:two_mtns}

\begin{figure}[htp]
	\centering
	\begin{subfigure}[b]{0.32\textwidth}
	\caption{$P_{\lambda, \tau}(t)$}
	\label{fig:two_mountains_example}
	\includegraphics[width=\textwidth]{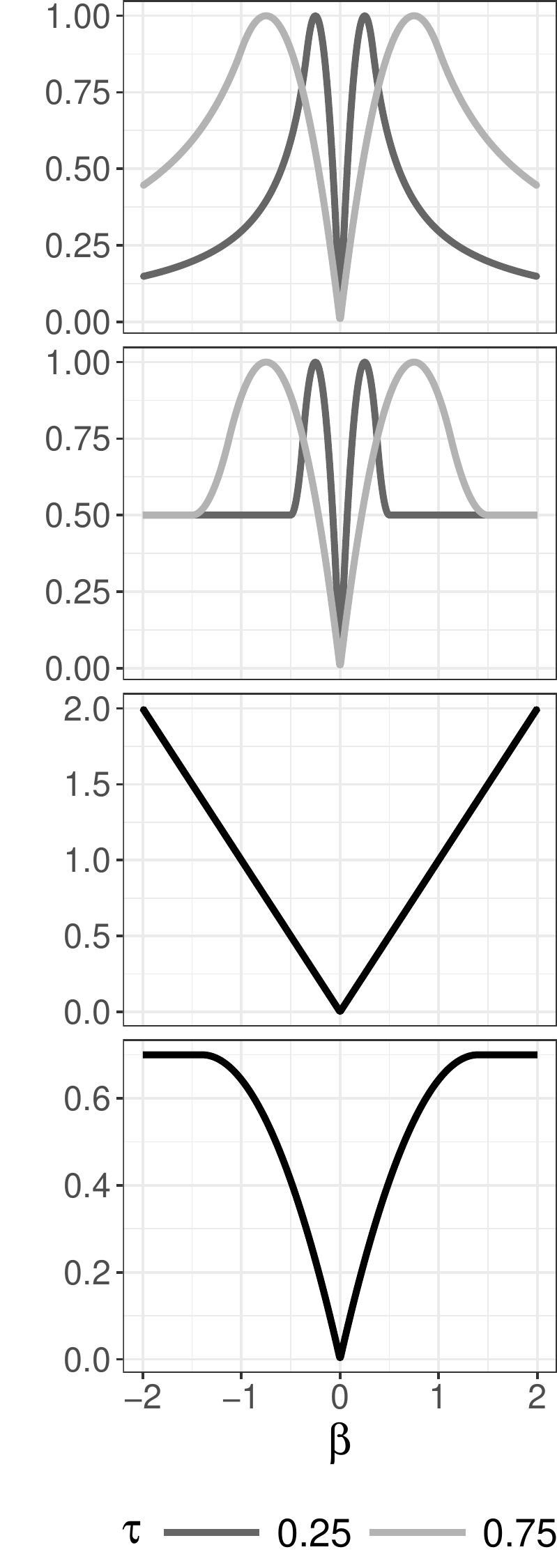}
	\end{subfigure}
	\begin{subfigure}[b]{0.32\textwidth}
	\caption{$P'_{\lambda, \tau}(t)$}
	\label{fig:pen_derivs}
	\includegraphics[width=\textwidth]{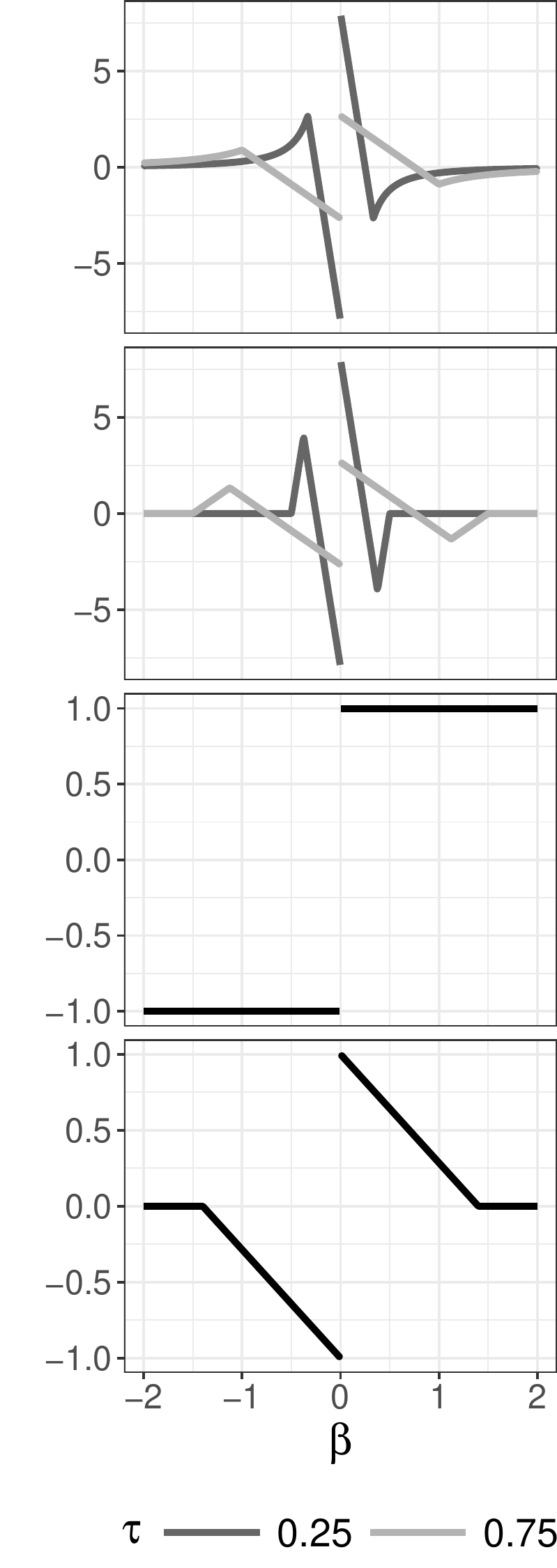}
	\end{subfigure}
	\begin{subfigure}[b]{0.32575\textwidth}
	\caption{$\argmin_{\beta} \frac{1}{2}(\beta - \theta)^2 + P_{\lambda, \tau}(|\theta|)$}
	\label{fig:two_mountains_threshold}
	\includegraphics[width=\textwidth]{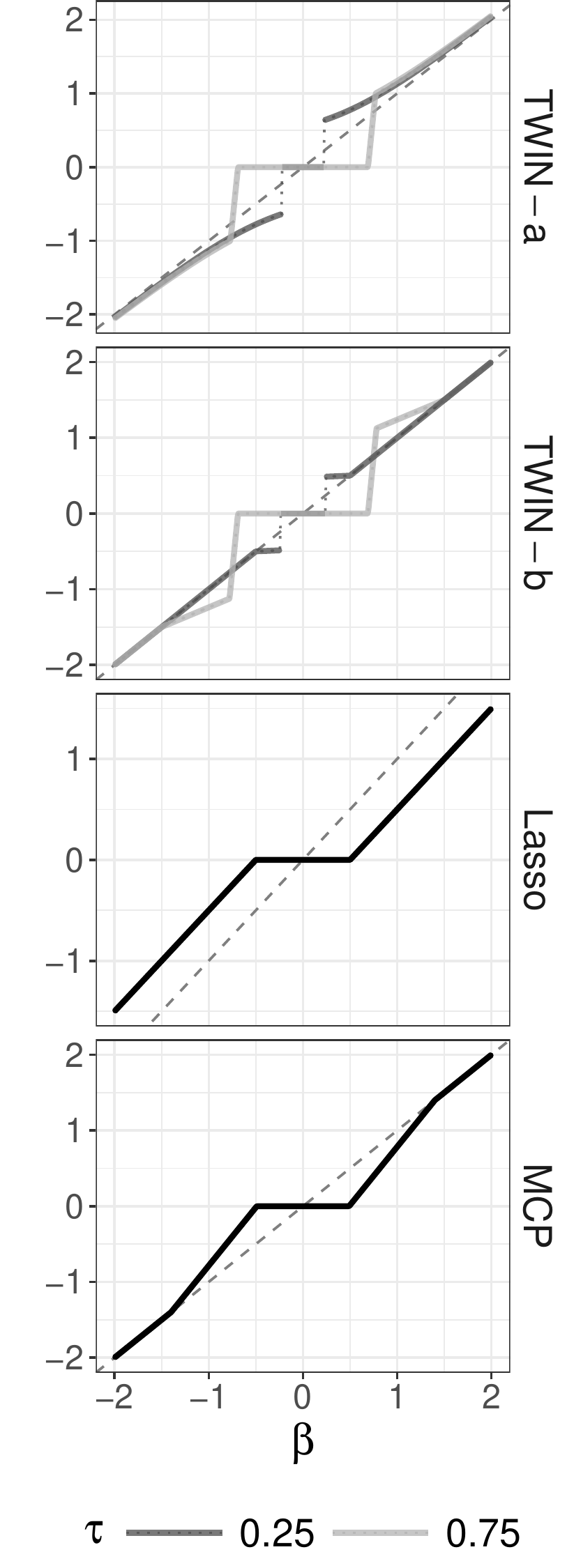}
	\end{subfigure}
	\caption{Panel (a) compares the penalty functions for TWIN-a and TWIN-b with the Lasso and MCP all with with $\lambda = 1$ (and $\lambda c = 1$ in the case of TWIN).  The extra tuning parameter $\gamma$ for MCP is set to 1.4. Panel (b) compares the corresponding derivative functions. Panel (c) compares the thresholding functions for all of the penalties. 
	}
	\label{fig:two_mountains_comparisons}
\end{figure}

The ``two mountains'' penalty class is defined by a general shape, which has the appearance of two mountains centered around the origin. Figure \ref{fig:two_mountains_example} depicts the archetypal shape of TWIN with two examples of the penalty class in comparison with the shapes of the Lasso penalty and the MCP.  The motivation of the two mountains shape is clear: it has a singularity at zero, thus allowing for variable selection, and it penalizes small coefficients more heavily and relaxes the  amount of penalization for large coefficients, effectuating the idea that variables with larger coefficients are more likely to be related to our response. Thus, it provides data-adaptive penalization of coefficients. However, the relationship between the magnitude of penalization is not monotone with coefficient size, as it is potentially unreasonable to assume that all small coefficients are necessarily unimportant. 

\begin{figure}[htp]
	\centering
	\includegraphics[width=1\textwidth]{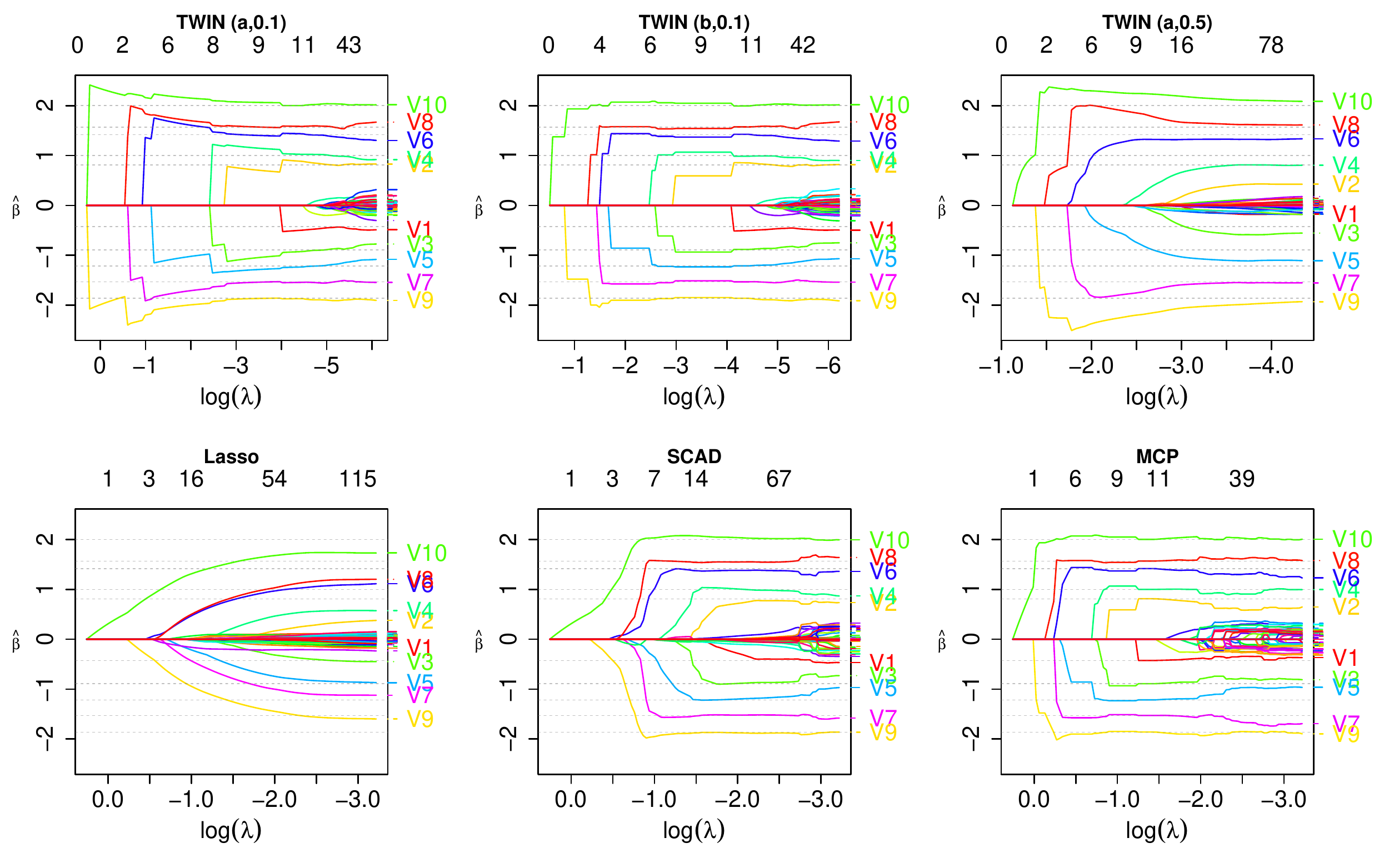}
	\caption{Plot of coefficient paths as the $\lambda$ tuning parameter is varied for TWIN-a and -b in comparison with  that of the Lasso, SCAD, and MCP. The top left plot is TWIN-a with $\tau=0.1$, the top middle is TWIN-b with $\tau=0.1$, and the top right is TWIN-a with $\tau=0.5$. Only variables $V1-V10$ have nonzero coefficients in this example and only these variables are labeled on the right of each plot if selected. }
	\label{fig:two_mountains_coef_paths}
\end{figure}

The TWIN penalty class $P_{\lambda,\tau}(t)$ is indexed by two parameters $\lambda,\tau>0$ and satisfies the following criteria:
\begin{enumerate}
	\item $P_{\lambda,\tau}(t)$ is continuous and nonnegative for $t \in \mathbb{R}^+$ with $P_{\lambda,\tau}(0) = 0$;
	\item $\sup_{\lambda>0}P_{\lambda,\tau}(t) = \infty$ for any $t\neq 0$;
	\item The derivative of the penalty is continuous except at the origin and  satisfies 
	\begin{itemize}
	\item $P'_{\lambda,\tau}(0+) = \lambda$, which enables the selection of variables,
	\item $P'_{\lambda,\tau}(t)$ is positive for $0<t<\tau$ and decreases to 0 such that $P'_{\lambda,\tau}(\tau)=0$,
	\item $P'_{\lambda,\tau}(t)$ is nonpositive for $t>\tau$, first decreasing in a neighborhood after $\tau$ and then increasing to 0, yielding a ``coefficient enlargement'' effect for a range of $t$ and (near) unbiasedness for large $t$,
	\end{itemize}
\end{enumerate}
When $P_{\lambda,\tau}$ is a member of the TWIN class, we call the minimizer of (\ref{eqn:penalized_least_squares}) a TWIN estimator. 
 Penalties that meet all of the two mountains (TWIN) criteria resemble two symmetrical hill or mountain shapes centered around 0 when taken as a function of $|t|$. 
The tuning parameter $\tau$ specifies the precise location of the peaks of the ``mountains'', i.e. where the penalty achieves its maximum value. The second criteria above guarantees that adjusting $\lambda$ will eventually result in a large enough penalty to set any coefficient to zero. The third property in criterion 3 above results in what we call coefficient enlargement in the sense that some estimates are slightly biased away from zero; see Figures \ref{fig:two_mountains_threshold} and \ref{fig:two_mountains_coef_paths}. The TWIN class can be further delineated based on the limiting behavior of $P_{\lambda,\tau}(t)$. The first subclass of TWIN penalties, which we call TWIN-a, is defined as all TWIN penalties which only achieve zero derivative in the limit. The second subclass, TWIN-b, has derivative equal to zero for all $t \ge d$ for some constant $d > 0$. This distinction results in different properties and our theoretical derivations will handle them separately.

The pattern of decreased penalization for $t>\tau$  is inspired by multiple testing procedures, wherein smaller $p$-values are compared with lower thresholds, for example \citet{benjamini1995controlling}. 
From the regression point of view (assuming equal variance of each coefficient estimate), smaller $p$-values correspond to stronger signals, i.e. variables with larger regression estimates. 
Thus the behavior of TWIN is opposite that of another recently proposed data-adaptive penalty, SLOPE \citep{bogdan2015}, which penalizes coefficients whose 
estimates are larger more heavily than those whose estimates are smaller.

In the following we introduce two specific TWIN penalties that will be used throughout this paper for demonstration purposes. While the theoretical results in this paper apply to all TWIN penalties, our numerical examples and our data analysis focus on the following two specific penalties in the TWIN class. 
\begin{example}[TWIN-a] \label{ex:pen_piecewise1} 
\begin{equation}\label{eqn:pen_piecewise1}
P_{\lambda, \tau}(t) = 
\begin{cases} 
    \lambda c(1 - (1 - t/\tau) ^ 2)  & t \leq m_1\tau \\
    \lambda c d_1\tau / t & t > m_1\tau
   \end{cases},
\end{equation}
where $d_1 > 0$ and $m_1 > 0$ are calculated such that the function above is continuous and has matching derivatives at $m_1$ and $c$ is a normalizing constant defined such that $P'_{\lambda, \tau}(0+)=\lambda$. The term $c$ can be dropped for clarity or ease of implementation. A direct calculation shows that $d_1 = 32/27$ and $m_1 = 4/3$. Note that letting $\tau\to 0$ and $\lambda\tau\to 1/(cd_1)$ yields $P_{\lambda,\tau}(t)=1/t$, which is the reciprocal Lasso of \citet{song2015}. 
\end{example}
\begin{example}[TWIN-b] \label{ex:pen_piecewise2}
	\begin{equation}\label{eqn:pen_piecewise2}
	P_{\lambda, \tau}(t) = 
	\begin{cases} 
	\lambda c(1 - (1 - t/\tau) ^ 2)  & t \leq m_2\tau \\
	\lambda c [(t - d_2) ^ 2 / \tau ^ 2 + h] & m_2\tau < t < d_2 \\
	\lambda c h & t > d_2
	\end{cases},
	\end{equation}
where $h \in (0,1)$ and $d_2 > 0$, $m_2 > 1$ are calculated such that the function above is continuous and has matching derivatives at $m_2\tau$ and $d_2$ and again $c$ is a normalizing constant defined such that $P'_{\lambda, \tau}(0+)=\lambda$. A straightforward calculation shows that $d_2 = (1 + \sqrt{2(1-h)})\tau$ and $m_2 = 1 + \sqrt{(1 - h)/2}$. The parameter $h$ can be chosen to balance  convexity of the penalty, and hence computational stability, with effect enlargement, however we simply choose $h = 1/2$. 
\end{example}
Examples 2.1 and 2.2 differ only in their behavior for $t > \tau$. 
\begin{remark}\label{remark:lasso_equivalence}
If $\tau\to\infty$ and $\lambda c / \tau \to \lambda^*/2$, both TWIN-a and TWIN-b become the Lasso penalty with tuning parameter $\lambda^*$.
\end{remark}

To better understand the behavior of TWIN penalties, let us consider the following univariate penalized least squares problem
\begin{equation}\label{eqn:univar_least_squares}
\frac{1}{2}(z - \theta)^2 + P_{\lambda, \tau}(|\theta|).
\end{equation}
\citet{fan2001} note that a good penalty function should meet three key criteria, namely i) (near) unbiasedness ii) sparsity, and iii) continuity of the minimizer of \eqref{eqn:univar_least_squares} with respect to $z$. TWIN meets the first two criteria, however, like for the hard-thresholding function \citep{antoniadis1997wavelets, fan1997comments} and for the reciprocal Lasso \citep{song2015}, it does not always meet the third. Specifically, for a range of values of $\tau$,  the minimizer of \eqref{eqn:univar_least_squares} is not continuous in $z$; see Figure \ref{fig:two_mountains_threshold}.  Thus, in some sense, the tuning parameter $\tau$ of TWIN offers a trade-off between continuity and computational stability. In spite of added computational instability, we find that TWIN with values of $\tau$ resulting in a discontinuous estimator often performs remarkably well in practice. Both examples TWIN-a (Example 2.1) and TWIN-b (Example 2.2) are computationally convenient, because they both admit closed-form solutions for univariate \eqref{eqn:univar_least_squares}, allowing for faster coordinate-descent algorithms with simple updates. 

Figure \ref{fig:two_mountains_coef_paths} displays the regularization paths of the Lasso, SCAD, MCP, TWIN-a and TWIN-b penalties from a simulated dataset with $n = 200$, $p = 1000$ among which only 10 active variables are related to the response, the covariates are generated independently from ${N}(0,\bsSigma)$ with $\bsSigma_{ij} = 0.5^{|i-j|}$, and $\bsz \sim{N}(0,I_n)$. The coefficients for the 10 active variables are given by $(-1/2, 2/3, -5/6, 1, -7/6, 4/3, -3/2, 5/3, -11/6, 2)$. The horizontal gray dashed lines are the oracle least squares estimates for the 10 active variables. Due to the low sample size, correlations between inactive variables and the response range between -0.21 and 0.22. The correlations between active variables and the response range in magnitude from 0.07 to 0.45 and are thus often dominated by random correlations with the response. Due in part to these correlations, the Lasso selects multiple inactive variables early on in the regularization path, a phenomenon studied rigorously in \citet{su2015}. 
Note that TWIN results in estimates which are inflated for a range of $\lambda$. Due to the fact that the derivative of the TWIN-a penalty is never exactly zero, it results in increased coefficient enlargement compared with TWIN-b. As we justify in Section \ref{sec:heuristics}, this added enlargement effect may be more beneficial in scenarios with strong correlations between covariates. Smaller coefficients, however, can still receive shrinkage towards zero by TWIN depending on the value of $\tau$. This behavior can be helpful in scenarios where prediction is a priority.

\subsection{Heuristics of TWIN}\label{sec:heuristics}

In this subsection, based on heuristic arguments, we provide insights into why the TWIN estimator yields reduced false discoveries compared with the Lasso, SCAD and MCP. The arguments in this section roughly follow and extend the arguments in \citet{su2015}.
For simplicity, in this section we fix $\sigma = 0$ as the following can be extended to cases with noise. Consider a Gaussian random design matrix $\bsX$ which has i.i.d. $N(0,1/n)$ entries and consider an oracle TWIN estimator with known true support $A^o = \{j:\beta_j\neq 0\}$ as obtained by
\begin{equation}
\label{eqn:reducedreg}
\widehat{\bsbeta}_{A^o} = \argmin_{\bsb_{A^o}\in\R^{\epsilon p}}\frac{1}{2}||\bsy - \bsX_{A^o}\bsb_{A^o}||^2 +\sum_{j\in {A^o}}P_{\lambda, \tau}(|b_j|),
\end{equation}
where $A^o$ is of approximate size $\epsilon p, 0<\epsilon<1,$ and $n,p\to\infty$. The matrix $\bsX_{A^o}$ is comprised of columns indexed by $A^o$ from the full design matrix $\bsX$. If $|\E_{\bsbeta_{A^o}}[\bsx_i'(\bsy - \bsX_{A^o}\widehat{\bsbeta}_{A^o})]|\leq \lambda$  for all $i\in \bar{A^o}$, where $\bar{A^o} = \{ 1, \dots, p \}\backslash A^o$, the KKT condition (\ref{eqn:kktcondnonconv}) suggests in expectation that extending $\widehat{\bsbeta}_{A^o}$ by adding zeros to $\bar{A^o}$ results in a solution of (\ref{eqn:penalized_least_squares}). If for some $j\in\bar{A^o}$, 
\begin{equation}
\label{eqn:constKKT}
|\E_{\bsbeta_{A^o}}[\bsx_j'(\bsy - \bsX_{A^o}\widehat{\bsbeta}_{A^o})]|>\lambda,
\end{equation}
then we must consider the reduced problem (\ref{eqn:reducedreg}) with support $A^o\cup\{j\}$ instead of $A^o$ in order to yield an equivalent solution with (\ref{eqn:penalized_least_squares}). 
Hence, (\ref{eqn:constKKT}) provides evidence of false discoveries.
Since $\widehat{\bsbeta}_{A^o}$ is independent of $\bsX_{\bar{A^o}}$, by conditioning on $\bsX_{A^o}$, $\E_{\bsbeta_{A^o}}[\bsx_j'(\bsy - \bsX_{A^o}\widehat{\bsbeta}_{A^o})]$ is normally distributed with mean zero and variance $n^{-1}\|\E_{\bsbeta_{A^o}}[\bsX_{A^o}(\bsbeta_{A^o} - \widehat{\bsbeta}_{A^o})]\|^2$.

To compare TWIN with the Lasso, observe that when $n>k$,  the largest singular value of $\bsX_{A^o}(\bsX_{A^o}'\bsX_{A^o})^{-1}$ is bounded, thus 
with probability approaching one,
\begin{equation}
\label{eqn:controlvarinexp}
\begin{aligned}
&  n^{-1}\|\E_{\bsbeta_{A^o}}[\bsX_{A^o}(\bsbeta_{A^o} - \widehat{\bsbeta}_{A^o})]\|^2 \\
= & n^{-1}\|\bsX_{A^o}(\bsX_{A^o}'\bsX_{A^o})^{-1}\E_{\bsbeta_{A^o}}[\text{sgn}(\widehat{\bsbeta}_{A^o})P'_{\lambda,\tau}(|\widehat{\bsbeta}_{A^o}|)]\|^2\\
 \leq  & c_0n^{-1}\left\{\lambda^2 \#\{j\in A^o, |\E_{\beta_j}[\widehat{\beta}_j]|<\gamma\lambda\} + \sup_{t\geq \gamma\lambda}|P'_{\lambda,\tau}(t)|^2 \#\{j\in A^o, |\E_{\beta_j}[\widehat{\beta}_j]|\geq\gamma\lambda\}\right\},
\end{aligned}
\end{equation}
where $c_0$ is some constant and  $\gamma$ is defined in (\ref{eqn:defofgamma}) which  indicates the region where $P'_{\lambda,\tau}$ is approximately  zero.
For Lasso estimators, we know $P'(\cdot)\equiv \lambda$ and thus the right-hand side of (\ref{eqn:controlvarinexp}) is of order $\lambda$ when $|A^o|$ is linear in $p$. In other words, Lasso estimators satisfy (\ref{eqn:constKKT}) for a number of variables in $\bar{A^o}$ linear in $p$, which causes a non-vanishing false discovery proportion; see \cite{su2015}. TWIN estimators, however, yield (near) unbiasedness, which results in $\sup_{t\geq \gamma\lambda}|P'_{\lambda,\tau}(t)|^2$ close to $0$. If the distribution of $\bsbeta_{A^o}$ is such that the minimal absolute value of true coefficients is larger than a certain threshold with a large probability (as in, e.g., \cite{tibshirani2011regression}), then $\#\{j\in A^o, |\E_{\beta_j}[\widehat{\beta}_j]|<\gamma\lambda\}/n \to 0$ and thus  the right-hand side of (\ref{eqn:controlvarinexp}) approaches $0$  for TWIN estimators, resulting in a vanishing proportion of false discoveries.

To compare TWIN with SCAD and MCP, we note that although these penalties are all (nearly) unbiased, TWIN penalties possess an \textit{enlargement} property for estimates with absolute values of a middling range; see,  Figure \ref{fig:two_mountains_comparisons} for illustration. The enlargement property can compensate in some sense  for the shrinkage error of estimates near zero. Specifically, we can bound the left-hand side of (\ref{eqn:controlvarinexp}) as follows:
\begin{equation}
\label{eqn:boundonsquaremcptwin}
\begin{aligned}
 n^{-1}\|\E_{\bsbeta_{A^o}}[\bsX_{A^o}(\bsbeta_{A^o} - \widehat{\bsbeta}_{A^o})]\|^2 & = n^{-1}\|\bsX_{A^o}(\bsX_{A^o}'\bsX_{A^o})^{-1}\E_{\bsbeta_{A^o}}[\text{sgn}(\widehat{\bsbeta}_{A^o})P'_{\lambda,\tau}(|\widehat{\bsbeta}_{A^o}|)]\|^2\\
&\leq c_1n^{-1}\|\E_{\bsbeta_{A^o}}[P'_{\lambda,\tau}(|\widehat{\bsbeta}_{A^o}|)]\|^2
\end{aligned}
\end{equation}
for some constant $c_1\geq 0$.
Since SCAD, MCP and TWIN yield shrinkage for weak signals, $P'(|\widehat{\beta}_j|)>0$ for small $\widehat{\beta}_j$. However, the enlargement property of TWIN enables 
$P'_{\lambda,\tau}(|\widehat{\beta}_j|)<0$ for $\beta_j$ with middling magnitudes, which compensates for positive $P'_{\lambda,\tau}(|\widehat{\beta}_j|)$'s and results in a smaller bound in (\ref{eqn:boundonsquaremcptwin}).
Thus for $j\in \bar{A^o}$, the conditional variance of  $\E_{\bsbeta_{A^o}}[\bsx_j'(\bsy - \bsX_{A^o}\widehat{\bsbeta}_{A^o})]$ has a smaller upper bound for TWIN, implying that TWIN is likely to give a smaller proportion of false discoveries than SCAD and MCP. 
Moreover, it is evident from extensive simulations in Section \ref{sec:simulation} that TWIN can be significantly better than SCAD and MCP in the linear sparsity regime with strong positive and negative correlations between inactive and active variables.  

%


\subsection{The role of the tuning parameter $\tau$}\label{sec:heuristics}

%
%

\begin{figure}[htp]
	\centering
	\includegraphics[width=1\textwidth]{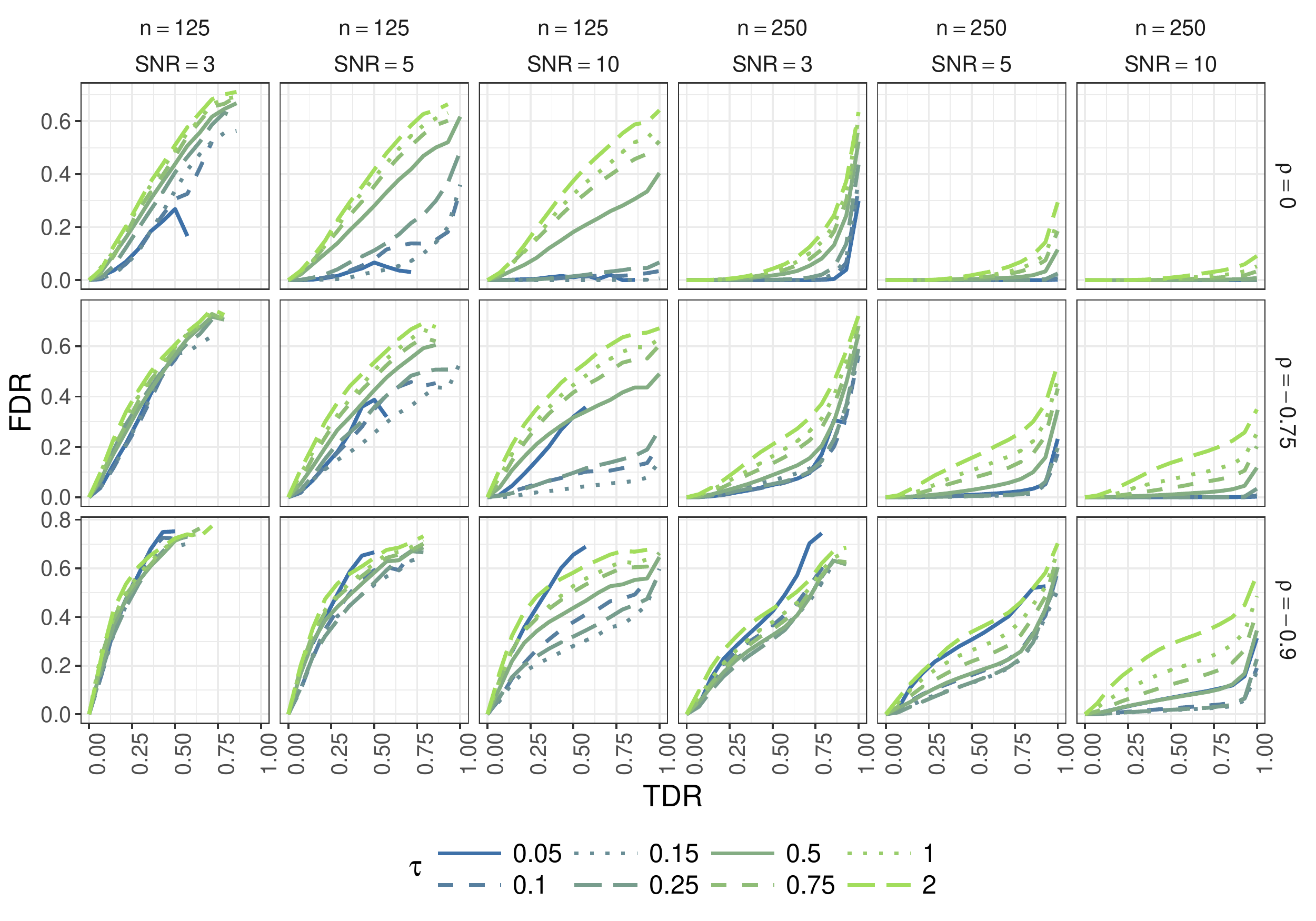}
	\caption{The results above are for a simulation with data generated under Model 3 described in Section \ref{sec:simulation}. Models are fit using the TWIN-a penalty.}
	\label{fig:sim_results_fdr_tpr_vary_tau_model1}
\end{figure}

TWIN's tuning parameter $\tau$ has an important impact on the selection behavior of TWIN.
We note that the reciprocal Lasso may yield overly sparse solutions when the underlying truth is not extremely sparse, and the Lasso may over-select variables when the underlying solution is indeed quite sparse. The tuning parameter $\tau$ balances between these two extremes. As $\tau$ tends to 0 and to $\infty$, TWIN becomes the reciprocal Lasso and the Lasso, respectively, allowing for a dynamic range of selection behavior. 
We now conduct a simulation study to investigate the finite sample properties of TWIN as $\tau$ is varied. 
Data are generated under model \eqref{eqn:linear_model} where the data-generating setup is described in Section \ref{sec:simulation} and the coefficients in the linear model are generated as described in Model 3 in Section \ref{sec:simulation}. We evaluate selection performance by investigating the average FDR versus true discovery rate of variable selection (TDR) curves as the tuning parameter $\lambda$ is varied. The curves are displayed in Figure \ref{fig:sim_results_fdr_tpr_vary_tau_model1}.

Generally, smaller values of $\tau$ tend to result in better selection characteristics as $\lambda$ is varied, however this comes at a cost of computational instability. The smallest value of $\tau$ considered works well in low correlation settings, but poorly with high correlations and when many covariates are selected.  Slightly larger values of $\tau$ such as 0.25 to 0.75 tend to have better performance in low signal settings with high correlations. Over all settings, including a more complete set of simulations presented in the Supplementary Material, values of 0.1 and 0.15 tend to work the best. However, in practice, it may be the case that  $\tau$ must be increased or decreased to some degree for ideal performance. 
In the Supplementary Material we further investigate the role of $\tau$ on prediction performance. The message is similar for prediction, however in scenarios with very low signal, larger values of $\tau$ are preferable if prediction is the primary goal. As $\tau$ increases, the model which minimizes the mean squared prediction error tends to be larger in size. 
 It is important to bear in mind that these investigations only span a small number of possibilities and thus do not always reflect how selection and estimation performance vary with $\tau$.


\section{Selection properties}
\label{sec:selectionconsis}

In this section we investigate the selection properties of TWIN estimators. In particular, we show that TWIN is selection consistent when a non-vanishing fraction of variables are important. Further, TWIN yields a finite sample FDR bound under orthogonal designs. We also provide universal values for both tuning parameters $\lambda$ and $\tau$ for which the selection consistency results hold. For low-dimensional regimes, these values do not depend on any unknown quantities other than the noise level.
We begin by studying the selection properties for orthogonal designs and then extend these results to random Gaussian designs.  Hereafter, we denote $\widehat{\bsbeta}$ as a TWIN estimator (distinctions between TWIN-a and TWIN-b will be made when warranted), $\bsbeta$ as the true coefficient vector, and 
\begin{equation}
\label{eqn:notationhataaok}
\widehat{A}\equiv\{j:\widehat{\beta}_j\neq 0\}, \quad A^o\equiv \{j:\beta_j\neq 0\},\quad\text{and}\quad k\equiv |A^o| = \#\{j:\beta_j\neq 0\}.
\end{equation}

\subsection{Orthogonal designs}

To gain insights about the TWIN estimator,  we first consider orthogonal designs. Under orthogonality, the optimality conditions for TWIN results in the following thresholding rule as the solution to
\begin{equation*}
\widehat{\bsbeta} = \text{sgn}(\bsX'\bsy)\left(|\bsX'\bsy| - P'_{\lambda,\tau}(|\widehat{\bsbeta}|)\right)_+,
\end{equation*}
where the sign function $\text{sgn}(t)\equiv I\{t>0\} - I\{t<0\}$. 
See Figure \ref{fig:two_mountains_threshold} for an illustration.
We note that when $|\widehat{\beta}_j|>\tau$, the absolute value of the resulting estimator is larger than the absolute value of the data. We call this effect the {enlargement} property since TWIN  amplifies estimates for moderately large $|\beta_j|$. However, TWIN yields (nearly) unbiased estimates for sufficiently large $|\beta_j|$. This overall behavior is different from the ``unbiasedness'' property of SCAD \citep{fan2001} and MCP \citep{zhang2010}, and is also different from the ``shrinkage'' property of the Lasso.  
We now present an upper bound of the FDR of TWIN under orthogonal designs.
\begin{proposition}
	\label{thm:FDRunderortho}
	Suppose that the data are generated from the linear model \eqref{eqn:linear_model} with an orthogonal design $\bsX$ and $\bsz\sim N(0,\sigma^2 I_p)$. Then for any $\alpha\in[0,1]$ the false discovery rate (FDR) and the family-wise error rate (FWER) for TWIN estimators obey,
	\begin{equation*}
	\text{FDR} = \E\left[\frac{\#\{j\in\widehat{A}\backslash A^o\}}{|\widehat{A}|\vee 1}\right] \leq \alpha\left(1-\frac{k}{p}\right),  \ \ \text{FWER} = \P\left\{\exists j\in \widehat{A}\backslash A^o\right\} = \alpha, 
	\end{equation*}
	by choosing 
	\begin{equation}
	\label{eqn:mintrabstpprimesigmphiinv}
	\min_{t\in\R}\{|t| + P'_{\lambda,\tau}(|t|)\} = \sigma\Phi^{-1}(1-\alpha/2p).
	\end{equation}
	If there are multiple pairs of $(\tau,\lambda)$ satisfying (\ref{eqn:mintrabstpprimesigmphiinv}), we select the pair resulting in the largest number of selected variables so as to increase power.
\end{proposition}
There are significant challenges in showing similar finite sample bounds for TWIN with a random design due to the estimation error of regression coefficients. See, for example, \citet{bogdan2015}.
Instead, we show that the FDR asymptotically approaches zero in 
Theorem \ref{thm:upperbdfalsepos}.

\subsection{Random designs}\label{sec:randomdesign_section}

In this section we study the selection properties of TWIN under random Gaussian designs where the columns of $\bsX$ have i.i.d. $N(0,1/n)$ entries so that the columns are approximately normalized. Random designs  are widely utilized in the statistics literature for studying regression methods.  See, for example, \cite{candes2006stable, zou2006adaptive, meinshausen2009lasso, van2009conditions, su2016}. Such designs are a sensible starting point for theoretical analysis of model selection properties due to weak correlations between the different predictors, as they obey restricted isometry properties \citep{candes2005decoding} or restricted eigenvalue conditions \citep{bickel2009simultaneous} with high probability. However, based on our numerical experiments, we suspect similar results may hold for designs with significant correlations and we leave this for future work. 

The rest of this section is organized as follows. We first introduce main assumptions in Section \ref{subsec:workingmodel} and then provide probability bounds of correct selection for TWIN in two cases: the global minimizer of (\ref{eqn:linear_model}) in the regular case where  $\text{rank}(\bsX)=p$ in  Section \ref{subsec:probboundsselection} and the local solution in the degenerate case where $\text{rank}(\bsX)<p$ in Section \ref{subsec:selechighdim}.

\subsubsection{Working assumptions and linear sparsity}
\label{subsec:workingmodel}

We assume throughout Section \ref{sec:randomdesign_section} that $p,n\to\infty$ and $n/p\to \delta$ for some constant $\delta > 0$. 
Further, as in \citet{su2015}, we assume that $\beta_1,\ldots,\beta_p$ are independent copies of a random variable $\Pi$ which satisfies $\E \Pi^2<\infty$ and $\P(\Pi\neq 0) = \epsilon$ where $\epsilon\in(0,1)$ is some constant. Hence, our assumptions accommodate linear sparsity where the expected value of $k$ equals to $\epsilon\cdot p$.
An asymptotic regime such as is discussed in  \cite{wainwright2009sharp}, among other works, where the proportion of nonzero coefficients vanishes in the limit of $p$ does not allow for linear sparsity.
As noted in \citet{su2015}, studying penalized regression methods in the linear sparsity regime yields theoretical results which accurately describe variable selection and estimation performance across a wide range of practical settings, as it can accommodate scenarios with relatively high dimension and a moderately low level of sparsity in addition to scenarios with very sparse signals. See \citet{bayati2012lasso, su2015} for extended discussion on the merits of the linear sparsity assumption.

For notational simplicity, we consider in Section \ref{sec:randomdesign_section} and Section \ref{sec:estimationproperties} that 
$\min_{t\in\R}\{|t| + P'_{\lambda,\tau}(|t|)\} = P'_{\lambda,\tau}(0+) = \lambda$, however the results in these two sections can be straightforwardly generalized to the case $0<\min_{t\in\R}\{|t| + P'_{\lambda,\tau}(|t|)\} < \lambda$. 
A  TWIN estimator $\widehat{\bsbeta}$ follows
\begin{equation}
\label{eqn:kktcondnonconv}
\begin{cases}
\bsx_j'(\bsy - \bsX\widehat{\bsbeta}) = \text{sgn}(\widehat{\beta}_j)P'_{\lambda,\tau}(|\widehat{\beta}_j|), & \widehat{\beta}_j\neq 0,\\
|\bsx_j'(\bsy - \bsX\widehat{\bsbeta})|\leq \lambda, & \widehat{\beta}_j =0.
\end{cases}
\end{equation}
Equations (\ref{eqn:kktcondnonconv}) are the Karush-Kuhn-Tucker (KKT) conditions for the global minimization of (\ref{eqn:penalized_least_squares}).
In general, solutions of (\ref{eqn:kktcondnonconv}) include all local minimizers of (\ref{eqn:penalized_least_squares}).


\subsubsection{Probability bounds for selection consistency}
\label{subsec:probboundsselection}

We first provide probability bounds for selection consistency when $n>p$ and $n$ and $p$ both tend to infinity. 
To clarify the distinction between TWIN-a and TWIN-b members of the TWIN class and to aid the presentation of theoretic results, we introduce an additional parameter $\gamma$ that describes the limiting behavior of $P'_{\lambda,\tau}(t)$ as follows:  
\begin{equation}
\label{eqn:defofgamma}
P'_{\lambda,\tau}(t)
\begin{cases} 
< 0 \text{ and } |P'_{\lambda,\tau}(t)| = o(\lambda), \quad  & \text{ when }t\geq \gamma\lambda, \text{ for  TWIN-a};\\
= 0, \quad  &\text{ when } t\geq \gamma\lambda,\text{ for  TWIN-b}.
\end{cases}
\end{equation}
In particular, TWIN-b becomes flat beyond a certain region while TWIN-a only has a 0 derivative beyond a certain range in the limit; see the illustration in Figure \ref{fig:pen_derivs}.
We consider the TWIN-a and TWIN-b variants of TWIN separately, as they exhibit slightly different behavior.  Recall that our theoretical exposition applies to all TWIN-a and TWIN-b penalties, not just the specific examples introduced in Section \ref{sec:two_mtns}.
We first present a non-asymptotic bound for selection consistency with TWIN-a penalties.
\begin{theorem}
\label{thm:lowdderivnegative}
Suppose that $n>p$, $\widehat{A}$ and $A^o$ are defined  in (\ref{eqn:notationhataaok}).
Let $\widehat{\bsbeta}$ be the TWIN-a estimator in (\ref{eqn:penalized_least_squares}) for  $\lambda\geq\{[(1-\vartheta)\sqrt{\delta/\epsilon}-1]^{-1}(1+\vartheta)+ 1\}(1+\vartheta)\sigma\sqrt{2\log p}$ and $\tau\geq(1-\delta^{-1/2} - \vartheta)^{-2}\lambda$ with any $\vartheta>0$. Then if $|\beta_j|> \gamma\lambda+\sigma\sqrt{(2+4\vartheta)\log k}(1 - \epsilon^{1/2}\delta^{-1/2}-\vartheta)^{-1}$ for all $j\in A^o$, we have
\begin{equation*}
\begin{aligned}
\P\left\{\widehat{A}\neq A^o\right\}& \leq \P\left\{\widehat{\bsbeta} \neq \widehat{\bsbeta}^o \text{ or } \text{sgn}(\widehat{\bsbeta})\neq \text{sgn}(\bsbeta)\right\}\\
& \leq e^{-n\sigma^2\vartheta^2/2}+e^{-k\vartheta^2/2}+3e^{-n\vartheta^2/2}+\sqrt{\pi\vartheta}k^{-\vartheta}.
\end{aligned}
\end{equation*}
In particular for large $n$, TWIN-a can arbitrarily control both type I and  type II errors to low levels under the linear sparsity regime, which yields $\P\{\widehat{A}= A^o\}\to 1$.
\end{theorem}
\begin{corollary}
\label{cor:twinanlargerthanp}
Suppose that $n>p$ and $\epsilon\leq 0.25$.
Let $\widehat{\bsbeta}$ be the TWIN-a estimator in (\ref{eqn:penalized_least_squares}) for  $\lambda_{a,\text{univ}}=(1+\delta^{-1/2})\sigma\sqrt{2\log p}$ and $\tau_{a,\text{univ}}=(0.99-\delta^{-1/2})^{-2}\lambda_{a,\text{univ}}$. Then if $|\beta_j|\geq \gamma\lambda_{a,\text{univ}}+\sigma\sqrt{2\log k}(1 - \epsilon^{1/2}\delta^{-1/2})^{-1}$ for all $j\in A^o$,  $\P\{\widehat{\bsbeta} \neq \widehat{\bsbeta}^o \text{ or } \text{sgn}(\widehat{\bsbeta})\neq \text{sgn}(\bsbeta)\}\to 0$.
\end{corollary}

The universal parameters $\lambda_{a,\text{univ}}$ and $\tau_{a,\text{univ}}$  do not require knowledge of the sparsity level. 
The condition $\epsilon\leq 0.25$ is only a technical requirement for the proof, however, it is 
a reasonable assumption in many applications. Now, we consider  the TWIN-b penalty and provide a similar non-asymptotic bound for its selection consistency.

\begin{theorem}
\label{thm:lowdimselectioncons}
Suppose that
$n>p$, $\widehat{A}$ and $A^o$ are defined in (\ref{eqn:notationhataaok}).
Let $\widehat{\bsbeta}$ be the TWIN-b estimator in (\ref{eqn:penalized_least_squares}) for  $\lambda\geq (1+3\vartheta)\sqrt{1-\epsilon\delta^{-1}}\sigma\sqrt{2\log p}$ and $\tau\geq(1-\delta^{-1/2} -\vartheta)^{-2}\lambda$ with any $\vartheta>0$. Then if $|\beta_j|> \gamma\lambda+\sigma\sqrt{(2+4\vartheta)\log k}(1 - \epsilon^{1/2}\delta^{-1/2} -\vartheta)^{-1}$  for all $j\in A^o$, we have
\begin{equation*}
\begin{aligned}
\P\left\{\widehat{A}\neq A^o\right\}& \leq \P\left\{\widehat{\bsbeta} \neq \widehat{\bsbeta}^o \text{ or } \text{sgn}(\widehat{\bsbeta})\neq \text{sgn}(\bsbeta)\right\}\\
& \leq e^{-\vartheta^2(n-k)\sigma^2/2}+ 2e^{-n\vartheta^2/2}+\sqrt{\pi\vartheta}(p-k)^{-\vartheta}+\sqrt{\pi\vartheta}k^{-\vartheta}.
\end{aligned}
\end{equation*}
In particular for large $n$, TWIN-b can arbitrarily control both type I and  type II errors to low levels under the linear sparsity regime, which yields
$\P\{\widehat{A}= A^o\}\to 1.$
\end{theorem}
\begin{corollary}
\label{cor:twinbnlargerthanp}
Suppose that $n>p$.
Let $\widehat{\bsbeta}$ be the TWIN-b estimator in (\ref{eqn:penalized_least_squares}) for  $\lambda_{b,\text{univ}}=\sigma\sqrt{2\log p}$ and $\tau_{b,\text{univ}}=(0.99-\delta^{-1/2})^{-2}\lambda_{b,\text{univ}}$. Then if $|\beta_j|\geq \gamma\lambda_{b,\text{univ}}+\sigma\sqrt{2\log k}(1 - \epsilon^{1/2}\delta^{-1/2})^{-1}$ for all $j\in A^o$, we have $\P\{\widehat{\bsbeta} \neq \widehat{\bsbeta}^o \text{ or } \text{sgn}(\widehat{\bsbeta})\neq \text{sgn}(\bsbeta)\}\to 0$.
\end{corollary}
Similar to Corollary \ref{cor:twinanlargerthanp},  the universal parameters $\lambda_{b,\text{univ}}$ and $\tau_{b,\text{univ}}$  do not require knowledge of the sparsity level. Extensive simulation studies demonstrating the effectiveness of the universal parameters and extended discussion on handling unknown noise level are presented in the Supplementary Material.

\subsubsection{Selection consistency for high-dimensional regression}
\label{subsec:selechighdim}

Now we consider the high-dimensional case where $p>n$ and $k<n$ and show the selection consistency of TWIN.
For brevity,  we only present results for TWIN-b as the following theorem can be generalized to the TWIN-a similarly as  Section \ref{subsec:probboundsselection}.

\begin{theorem}
\label{thm:upperbdfalsepos}
Suppose that $p> n$, $\widehat{A}$ and $A^o$ are defined in (\ref{eqn:notationhataaok}). 
Let $\widehat{\bsbeta}$ be the TWIN-b estimator  in (\ref{eqn:penalized_least_squares}) for $\lambda\geq \max\{(1+3\vartheta)\sqrt{1-\epsilon\delta^{-1}}\sigma\sqrt{2\log p}, 2[1+\vartheta+\sqrt{(\epsilon/\delta+1)/2}]\sigma\sqrt{2\tilde{c}+1}\}$ and $\tau\geq(1- \sqrt{(\epsilon/\delta+1)/2}-\vartheta)^{-2}\lambda$ with any $\vartheta>0$ and  $\tilde{c} \equiv [(1-\epsilon)\log(1-\epsilon) - (\delta-\epsilon)\log(\delta-\epsilon) - (1-\delta)\log(1-\delta)]/\delta$.
Then if $|\beta_j| >\gamma\lambda+\sigma\sqrt{(2+4\vartheta)\log k}(1 - \epsilon^{1/2}\delta^{-1/2} -\vartheta)^{-1}$ for all $j\in A^o$ and $\epsilon/\delta\leq 0.12$,  we have
\begin{equation*}
\begin{aligned}
\P\left\{\widehat{A}\neq A^o\right\} & \leq \P\left\{\widehat{\bsbeta} \neq \widehat{\bsbeta}^o \text{ or } \text{sgn}(\widehat{\bsbeta})\neq \text{sgn}(\bsbeta)\right\}\\
& \leq e^{-\vartheta^2(n-k)\sigma^2/2}+ 2e^{-n\vartheta^2/2}+\sqrt{\pi\vartheta}(p-k)^{-\vartheta}\\
& \quad+\sqrt{\pi\vartheta}k^{-\vartheta}+ \left\{[\tilde{c}+(n-k)^{-1}]\sqrt{2\pi(n-k)}\right\}^{-1}.
\end{aligned}
\end{equation*}
\end{theorem}

\begin{corollary}
\label{cor:upperbdfalsepos}
Suppose that $p> n$. 
Let $\widehat{\bsbeta}$ be the TWIN-b estimator  in (\ref{eqn:penalized_least_squares}) for $\lambda_{b,\text{univ}}  = \sigma\sqrt{2\log p}$ and $\tau'_{\text{univ}}\geq[0.99-\sqrt{(\epsilon/\delta+1)/2}]^{-2}\lambda_{b,\text{univ}}$.
Then if $|\beta_j| \geq\gamma\lambda_{b,\text{univ}} + \sigma\sqrt{2\log k}(1-\epsilon^{1/2}\delta^{-1/2})^{-1}$ for all $j\in A^o$ and $\epsilon/\delta\leq 0.12$,  we have $\P\{\widehat{\bsbeta} \neq \widehat{\bsbeta}^o \text{ or } \text{sgn}(\widehat{\bsbeta})\neq \text{sgn}(\bsbeta)\}\to 0$.
\end{corollary}

The  parameter $\lambda_{b,\text{univ}}$ is the same as in Corollary \ref{cor:twinbnlargerthanp} and does not require knowledge about the sparsity level. For $\tau'_{\text{univ}}$ to avoid a requirement of exact knowledge of the sparsity level, we can use  a  prior upper bound on $\epsilon$, denoted by $\epsilon'$, and set $\tau'_{\text{univ}}=[0.99-\sqrt{(\epsilon'/\delta+1)/2}]^{-2}\lambda_{b,\text{univ}}$, which satisfies the condition of Corollary \ref{cor:upperbdfalsepos}.

Theorem \ref{thm:upperbdfalsepos} and Corollary \ref{cor:upperbdfalsepos} show that in the case of high-dimensionality and linear sparsity, TWIN estimators have false discovery rate and true discovery rate (TDR) obeying
\begin{equation*}
\lim_{n\to\infty}\text{FDR} = \lim_{n\to\infty}\E\left[\frac{\#\{j\in\widehat{A}\backslash A^o\}}{|\widehat{A}|\vee 1}\right] = 0, \ \ \lim_{n\to\infty}\text{TDR} = \lim_{n\to\infty}\E\left[\frac{\#\{j\in\widehat{A}\cap A^o\}}{k\vee 1}\right] = 1.
\end{equation*}
Theorem \ref{thm:upperbdfalsepos} also implies that $n= (\delta/\epsilon+o(1))k>8.33k$ is sufficient for perfect recovery. It is known in the compressed sensing literature that in the  no noise case, $n$ Gaussian samples with $n\geq 2(1+o(1))k\log(p/k) = 2(1+o(1))k\log(1/\epsilon)$  are required for perfect support recovery using $l_1$-based methods; see, e.g., \cite{donoho2010exponential}. Stricter conditions are usually assumed in the statistics literature for perfect recovery, for example, $k/p\to0$ in \cite{song2015} and 
$(k\log p)/n\to0$ in \cite{su2016}.


\section{Estimation properties}
\label{sec:estimationproperties}

In this section, we investigate the minimax optimality of estimation with TWIN estimators under random Gaussian designs and linear sparsity. In the Supplementary Material, we present corresponding results for minimax optimality under orthogonal designs.
%
%
As noted in the literature \citep{su2016}, minimax optimality results for orthogonal designs do not in general imply similar results for Gaussian designs because of the sample correlations among the columns of Gaussian designs.
The goal of this section is to establish the minimax optimality of TWIN estimators under Gaussian  designs and linear sparsity.

\subsection{Risk lower bound under linear sparsity}
The following result gives an explicit lower bound of asymptotic risk under the linear sparsity and random Gaussian designs. 
\begin{theorem}
\label{thm:randomlowrbdesterror}
Suppose that  $k/p\to \epsilon\in(0,1)$ as $p\to\infty$. Let $\bsbeta$ be from the model (\ref{eqn:linear_model}) and  the columns of $\bsX$ have i.i.d. $N(0,1/n)$ entries. Then for any $\vartheta\in(0,1)$, we have
\begin{equation*}
\inf_{\widetilde{\bsbeta}}\sup_{\|\bsbeta\|_0\leq k}\P\left\{\frac{\|\widetilde{\bsbeta} - \bsbeta\|^2}{2\sigma^2k\log(1/\epsilon)}>1-\vartheta\right\} =1,
\end{equation*}
where the infimum is taken over all measurable estimators. 
\end{theorem}
Similar results for random designs can be found in the literature; see, for example, \cite{ye2010rate, raskutti2011minimax, su2016}. However, the main difference of such results and Theorem \ref{thm:randomlowrbdesterror} is that   instead of assuming  $k/p\to 0$ and $(k\log p)/n\to0$, Theorem \ref{thm:randomlowrbdesterror} considers the linear sparsity regime $k/p\to\epsilon$ with unknown constant $\epsilon\in(0,1)$ and provides the exact constant in front of the rate.

\subsection{Risk upper bounds for TWIN estimators}

We first give a probabilistic bound on the asymptotic risk for TWIN-a estimators.

\begin{theorem}
\label{thm:guassupperbdesterror}
Suppose that $p,n\to\infty$ with $n/p\to\delta$ for some constant $\delta>1$ and $k/p\to\epsilon$ for some constant $0<\epsilon<1$. Let $\widehat{\bsbeta}$ be the TWIN-a estimator in (\ref{eqn:penalized_least_squares}) for $\lambda = \{[(1-\vartheta)\sqrt{\delta/\epsilon}-1]^{-1}(1+\vartheta)+ 1\}(1+\vartheta)\sigma\sqrt{2\log p}$ and $\tau\geq(1-\delta^{-1/2}-\vartheta)^{-2}\lambda$ for arbitrary $\vartheta>0$. Then, 
\begin{equation}
\label{eqn:guassupperbdesterror}
\sup_{\|\bsbeta\|_0\leq k}\P\left\{\frac{ \| \bsbeta- \widehat{\bsbeta}\|^2}{C_1(\epsilon,\delta)\cdot 2\sigma^2k\log p}\leq 1\right\}\to 1,
\end{equation}
where the constant $C_1(\epsilon,\delta) = \left\{\frac{\sqrt{3}}{[(1-\lambda\tau^{-1})\delta^{1/2}\epsilon^{-1/2}-2]^{1/2}}+1\right\}^2[(\delta^{1/2}\epsilon^{-1/2}-1)^{-1}+1]^2$. 
\end{theorem}

We make the following remarks on the above theorem.  
First, comparing Theorem \ref{thm:guassupperbdesterror} with the lower bound result Theorem \ref{thm:randomlowrbdesterror}, there is a difference in their logarithm terms, which is actually due to the unknown sparsity level. More specifically, when $k$ is unknown, a tight upper bound for $1/\epsilon$ is $p$. Hence, TWIN-a estimators are minimax rate optimal. 
Second, $C_1(\epsilon,\delta)$ is close to one when $\epsilon$ is small, which meets the constant in Theorem \ref{thm:randomlowrbdesterror}. Third,  we have shown in Corollary \ref{cor:twinanlargerthanp} that universal tuning parameters $\lambda_{a,\text{univ}}$ and $\tau_{a,\text{univ}}$  yield selection consistency. The following result shows further that these universal tuning parameters yield asymptotic estimation risk  with the minimax optimal rate.
\begin{corollary}
\label{cor:estimationoftwinacor}
Suppose that $p,n\to\infty$ with $n/p\to\delta$ for some constant $\delta>1$ and $k/p\to\epsilon$ for some constant $0<\epsilon\leq 0.25$. Let $\widehat{\bsbeta}$ be the TWIN-a estimator in (\ref{eqn:penalized_least_squares}) for $\lambda_{a,\text{univ}} = (1+\delta^{-1/2})\sigma\sqrt{2\log p}$ and $\tau_{a,\text{univ}}=(0.99-\delta^{-1/2})^{-2}\lambda_{a,\text{univ}} $. Then, 
$\sup_{\|\bsbeta\|_0\leq k}\P\{\|\bsbeta - \widehat{\bsbeta}\|^2/[C'_1(\epsilon,\delta)\cdot 2\sigma^2k\log p]\leq 1\}\to 1$ with constant $C'_1(\epsilon,\delta) = \{\sqrt{3}/[(1.98-\delta^{-1/2})\epsilon^{-1/2}-2]^{1/2}+1\}^2(1+\delta^{-1/2})^2$. 
\end{corollary}

A similar probabilistic bound on the asymptotic risk holds for TWIN-b estimators.
\begin{theorem}
\label{thm:guassupperbdesterrorb}
Suppose that $p,n\to\infty$ with $n/p\to\delta$ for some constant $\delta>1$ and $k/p\to\epsilon$ for some constant $0<\epsilon<1$. Let $\widehat{\bsbeta}$ be the TWIN-b estimator in (\ref{eqn:penalized_least_squares}) for $\lambda = (1+3\vartheta)\sqrt{1-\epsilon\delta^{-1}}\sigma\sqrt{2\log p}$ and $\tau\geq(1-\delta^{-1/2}-\vartheta)^{-2}\lambda$ for arbitrary $\vartheta>0$. Then, 
\begin{equation}
\sup_{\|\bsbeta\|_0\leq k}\P\left\{\frac{ \|\bsbeta - \widehat{\bsbeta}\|^2}{C_2(\epsilon,\delta)\cdot 2\sigma^2k\log p}\leq 1\right\}\to 1,
\end{equation}
where the constant $C_2(\epsilon,\delta) = \left[\frac{\sqrt{3}}{(\delta^{1/2}\epsilon^{-1/2}-2)^{1/2}}+1\right]^2(1-\epsilon\delta^{-1})$. 
\end{theorem}
We make the following remarks on the above theorem.  First, similar to the discussion after Theorem \ref{thm:guassupperbdesterror}, TWIN-b estimators are minimax rate optimal. Second, 
$C_2(\epsilon,\delta)$ is close to one when $\epsilon$ is small, which also meets the constant in Theorem \ref{thm:randomlowrbdesterror}. Third, we note $C_1(\epsilon,\delta)>C_2(\epsilon,\delta)$, which implies TWIN-b estimators achieve a smaller upper bound of asymptotic risk than TWIN-a estimators when $\epsilon > 0$. Heuristically, this is due to the unbiasedness property of the TWIN-b estimators, whereas TWIN-a estimators are only nearly unbiased and often result in stronger enlargement effects.
Fourth,   Corollary \ref{cor:twinbnlargerthanp} shows that universal tuning parameters $\lambda_{b,\text{univ}}$ and $\tau_{b,\text{univ}}$  yield selection consistency and now the following result shows they also yield the minimax optimal rate. 
\begin{corollary}
\label{cor:estimationoftwinbcor}
Suppose that $p,n\to\infty$ with $n/p\to\delta$ for some constant $\delta>1$ and $k/p\to\epsilon$ for some constant $0<\epsilon<1$. Let $\widehat{\bsbeta}$ be the TWIN-b estimator in (\ref{eqn:penalized_least_squares}) for $\lambda_{b,\text{univ}} = \sigma\sqrt{2\log p}$ and $\tau_{b,\text{univ}}=(0.99-\delta^{-1/2})^{-2}\lambda_{b,\text{univ}}$. Then, 
$\sup_{\|\bsbeta\|_0\leq k}\P\{\|\bsbeta - \widehat{\bsbeta}\|^2/[C'_2(\epsilon,\delta)\cdot 2\sigma^2k\log p]\leq 1\}\to 1$ with constant $C'_2(\epsilon,\delta) = [\sqrt{3}/(\delta^{1/2}\epsilon^{-1/2}-2)^{1/2}+1]^2$. 
\end{corollary}

Finally, we remark that results in Theorem \ref{thm:guassupperbdesterror} and  \ref{thm:guassupperbdesterrorb} can be generalized to the high-dimensional case where $p> n$ and $k<n$ as in Section \ref{subsec:selechighdim}.

\section{Numerical studies}
\label{sec:simulation}

In this section we seek to demonstrate the variable selection properties of the TWIN penalty under various challenging and realistic high dimensional scenarios. In this section we simulate data under model (\ref{eqn:linear_model}) where the number of non-zero elements in $\bsbeta$ is very small relative to the dimension $p$. We generate $\bsX$ from a multivariate Gaussian distribution with covariance matrix  $\bsSigma \in \R^{p\times p}$ with $\bsSigma_{ij} = \rho^{|i-j|}$.
Larger $|\rho|$ indicates stronger correlations between predictors. The correlation parameter $\rho$ is varied from $(0, -0.75, -0.90)$, the sample size is set to $125$ and $250$, and $p$ is set to 1000. We focus on $\rho \leq 0$, as most data for regression tasks exhibit both positive and negative correlations. We set the variance of the error term  such that the signal-to-noise ratio (SNR), defined as $SNR = {\sqrt{\bsbeta^T\bsSigma\bsbeta}}/{\sigma}$,
is 3, 5, and 10. Given the number of active variables in the models considered below, this range of the signal-to-noise ratio makes it very difficult to recover the active variables.
In all of the above settings the $k$ active coefficients are chosen uniformly at random from all $p$ covariates with magnitudes of the active coefficients generated under the following two schemes: i.) independent random variates from a uniform distribution on $[-2, 0.5] \cup [0.5, 2]$ and ii.) $(-c)^{j-1}$ for the $j$th of $k$ active variables. Under Models 1 and 3, we generate coefficients from scheme i.) with $k=50$ and $k=25$, respectively, and under Models 2 and 4 we generate coefficients from scheme ii.) with $c=0.95$ and $c=0.8$, respectively, and $k=50$ and $k=25$, respectively.
%
%
%
%
%
%
%
%
%
%
%
The beta-min condition is not satisfied under scheme ii.), as the smallest nonzero coefficients are close to 0 and much smaller than the largest coefficients, whereas under scheme i.) coefficients are bounded away from 0.

We compare TWIN-a and TWIN-b with  the Lasso, SCAD, and MCP. We use the \texttt{R} package \texttt{ncvreg} \citep{breheny2011coordinate} to implement SCAD and MCP and use the \texttt{R} package \texttt{glmnet} \citep{friedman2010regularization} to implement the Lasso.	 Throughout the simulations, we set the $\gamma$ tuning parameter for MCP to be 1.4 as recommended in \cite{zhang2010} and for SCAD to be 3.7, as recommended in \citet{fan2001}. The bandwidth tuning parameter $\tau$ of TWIN-a  and TWIN-b is set to be 0.1 throughout the simulations. In the Supplementary Material we introduce two algorithms for computation for the TWIN penalty. The first algorithm is a  modification of coordinate descent and is denoted as CD and the second algorithm is a hybrid local linear approximation \citep{zou2008one} and coordinate descent algorithm, which we denote as MCLLA for mixed coordinate local linear approximation. We investigate the performance of TWIN using both CD and MCLLA using random coordinate updates instead of cyclical updates, as described in the Supplementary Material.

\begin{figure}[htp]
	\centering
	\includegraphics[width=0.9\textwidth]{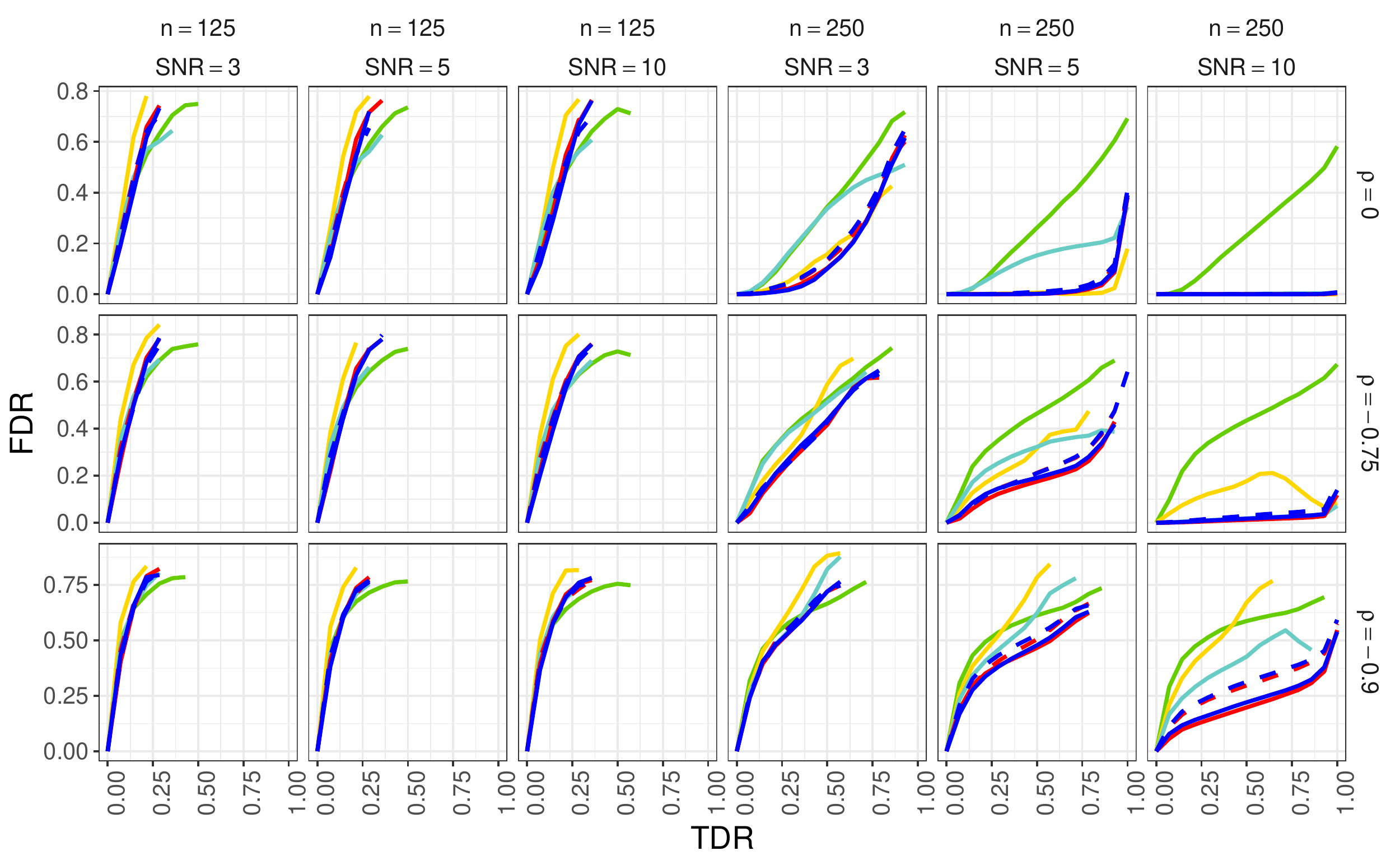}
	\includegraphics[width=0.9\textwidth]{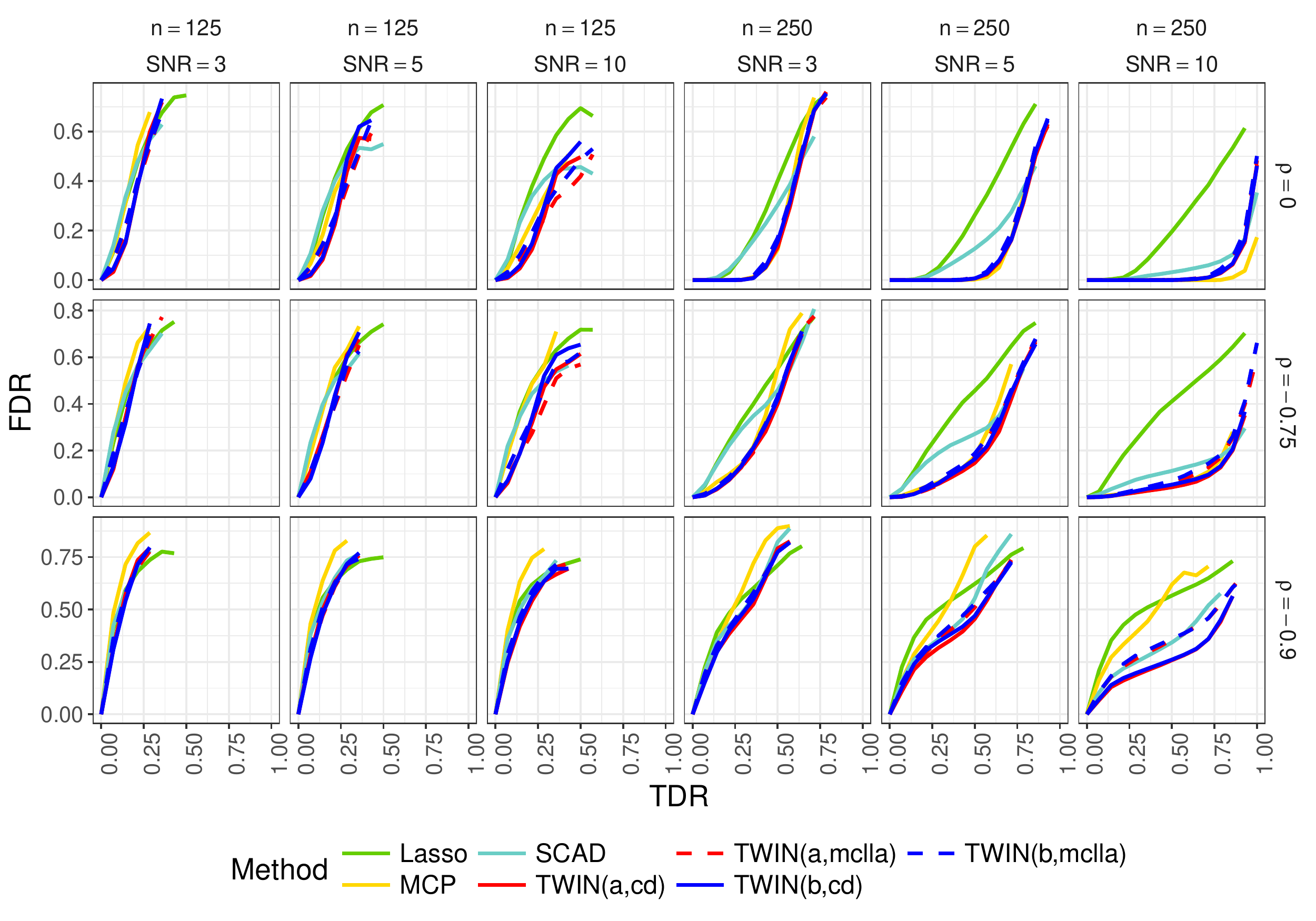}
	\caption{The results above are for a simulation with data generated under Model 1 (top panel) and Model 2 (bottom panel).}
	\label{fig:sim_results_fdr_tpr_model1}
\end{figure}

\begin{figure}[htp]
	\centering
	\includegraphics[width=0.9\textwidth]{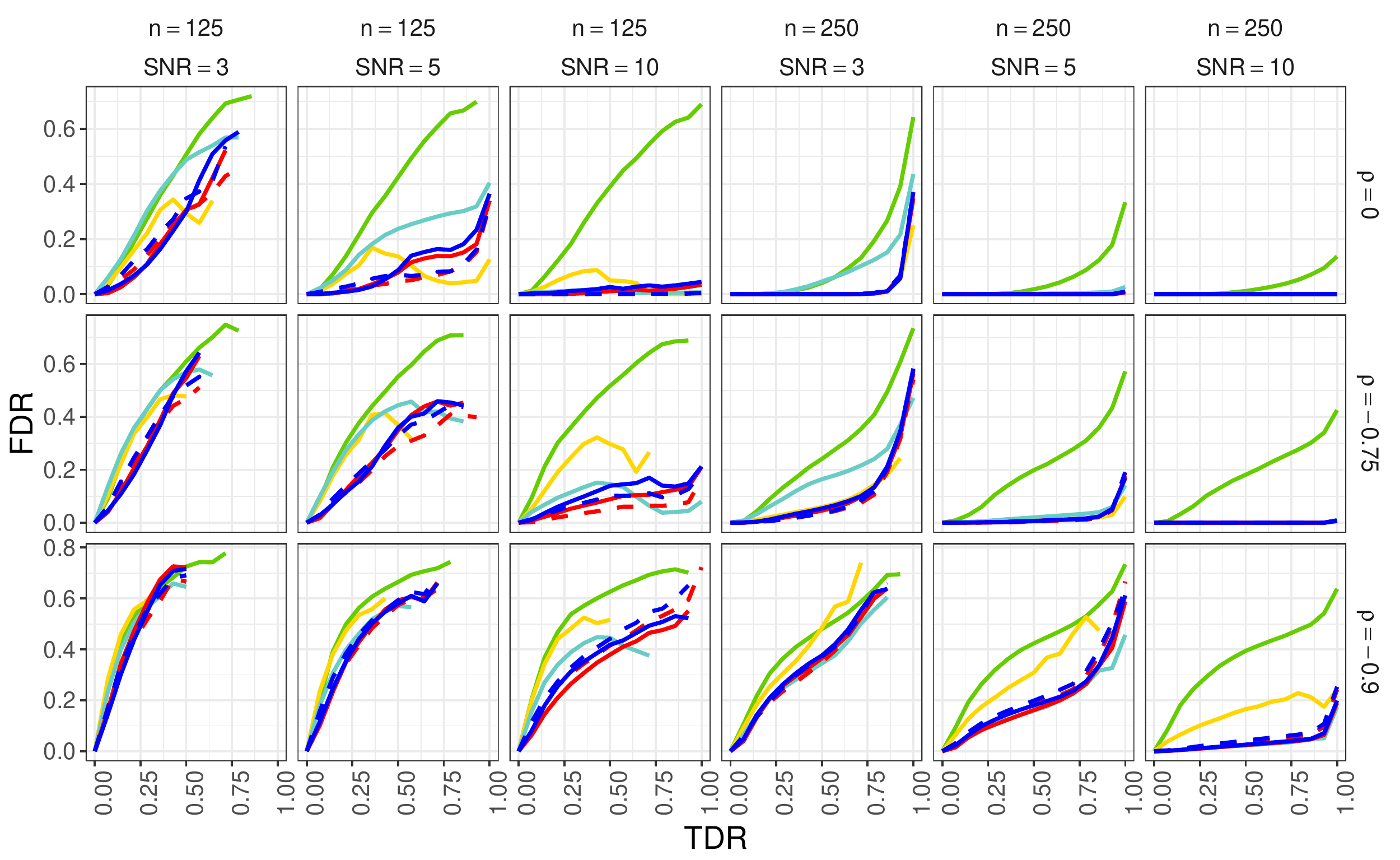}
	\includegraphics[width=0.9\textwidth]{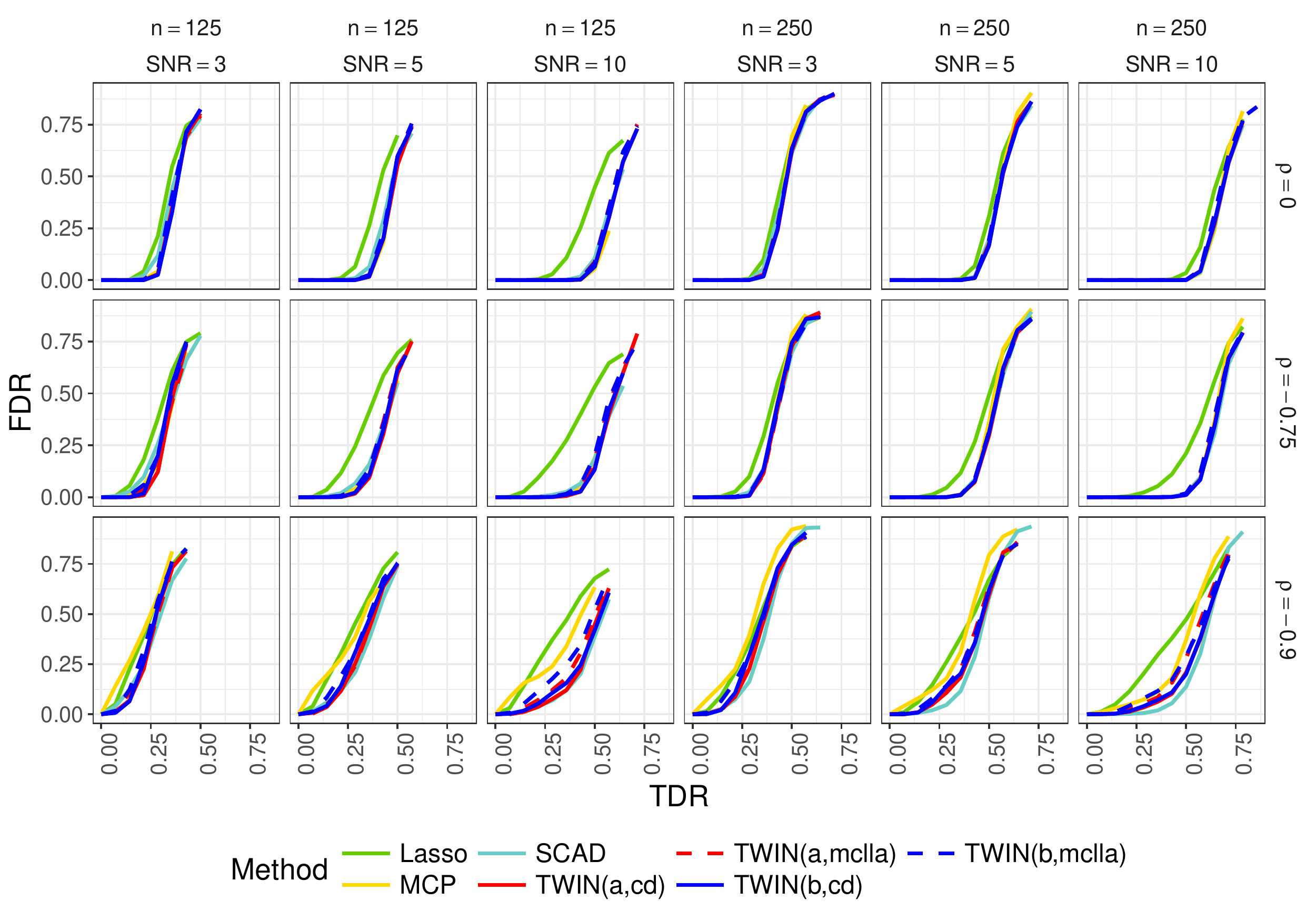}
	\caption{The results above are for a simulation with data generated under Model 3 (top panel) and Model 4 (bottom panel).}
	\label{fig:sim_results_fdr_tpr_model3}
\end{figure}

\begin{figure}[htp]
	\centering
	\includegraphics[width=0.9\textwidth]{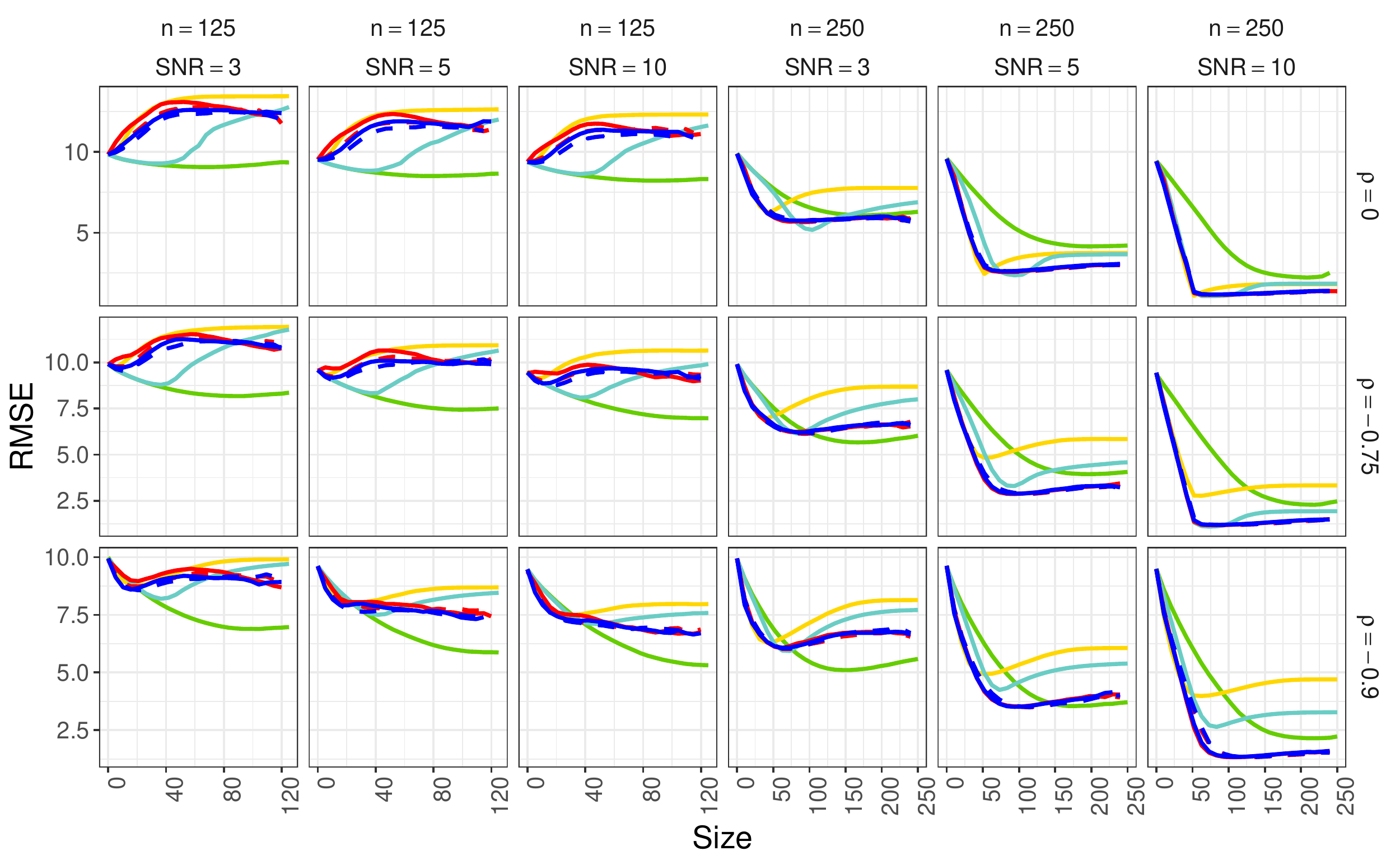}
	\includegraphics[width=0.9\textwidth]{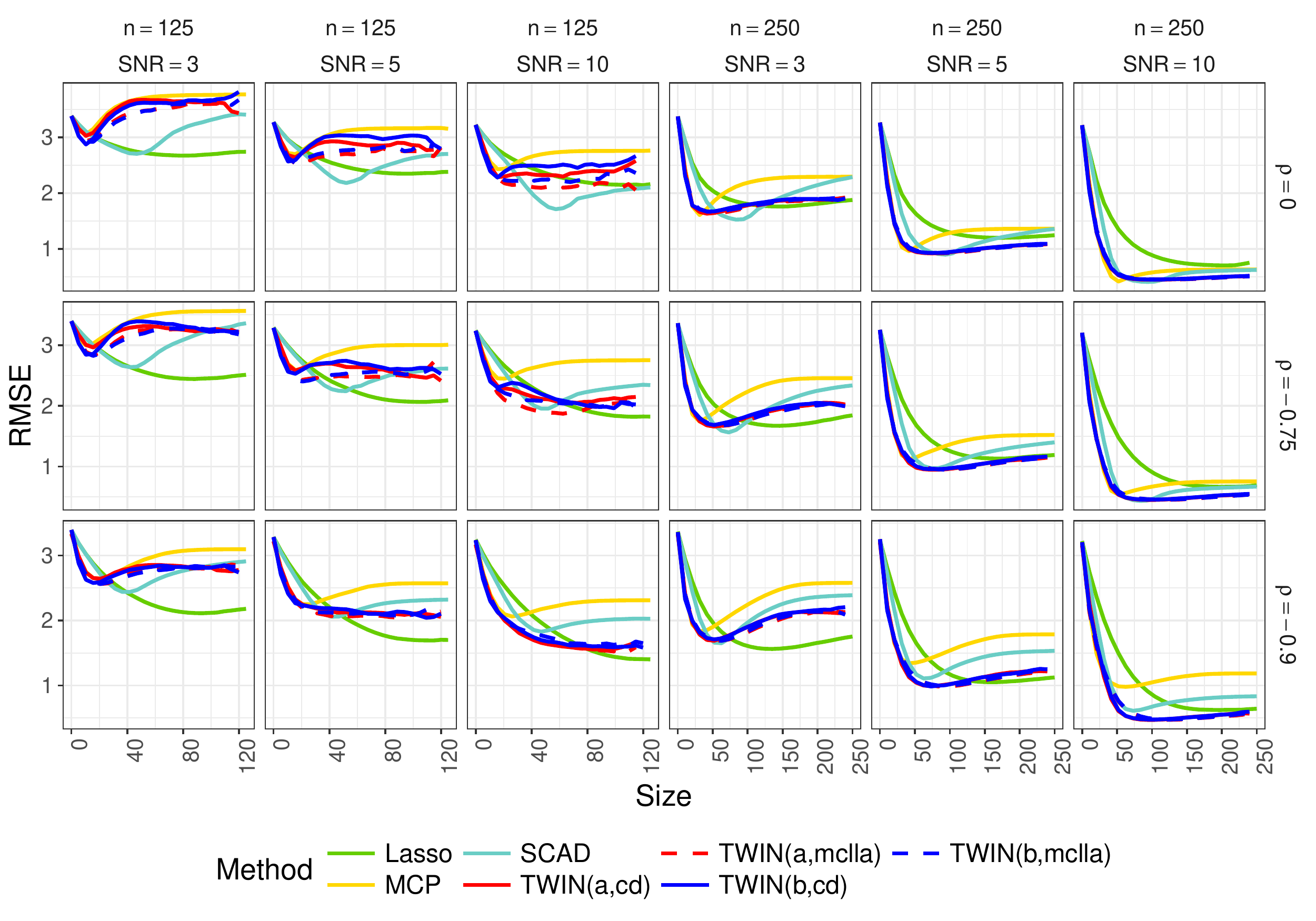}
	\caption{The results above are for a simulation with data generated under Model 1 (top panel) and Model 2 (bottom panel).}
	\label{fig:sim_results_rmse_size_model1}
\end{figure}

As we wish to understand the underlying operating characteristics of all methods with respect to FDR and TDR, we evaluate each method by investigating the relationship between FDR and TDR as the selection tuning parameter $\lambda$ is varied.  In Figures \ref{fig:sim_results_fdr_tpr_model1} and \ref{fig:sim_results_fdr_tpr_model3} we display average FDR versus TDR curves under Models 1-4 averaged over 100 independent datasets. To demonstrate predictive performance under the same simulation settings, we display in Figure \ref{fig:sim_results_rmse_size_model1} the average square root of the mean squared prediction error (RMSE) versus the number of nonzero coefficients for each method. Due to space concerns, prediction results for Models 3 and 4 are included in the Supplementary Material. The RMSE is evaluated on an independent dataset of size 5000. The independent dataset is generated anew for each replication of the simulation study.

We first evaluate the variable selection results. In settings with more active variables (Models 1 and 2), both TWIN-a and TWIN-b outperform all other methods when there are correlations between covariates. In the no correlation setting ($\rho=0$), TWIN-a and TWIN-b both outperform SCAD and the Lasso, but have similar albeit slightly worse performance than MCP in high SNR and/or sample size settings. However, TWIN-a and TWIN-b tend to perform better than MCP in most low-signal and/or low sample size settings. In settings with 25 active variables (Models 3 and 4), the comparisons are similar, except SCAD performs nearly as well as TWIN-a and TWIN-b when $\rho=-0.9$ under Model 3 and slightly better than TWIN-a and TWIN-b when $\rho=-0.9$ under Model 4. 

Regarding prediction performance, we first consider results under Models 1 and 2. In low SNR settings, the Lasso and SCAD tend to perform the best, with the Lasso achieving the smallest minimum RMSE, albeit with models which are on average much larger than models which minimize RMSE under different penalties. Like MCP and unlike SCAD and the Lasso, both TWIN-a and TWIN-b tend to achieve their minimum RMSE with models that are of approximately the correct size of the underlying data-generating model. In high correlation settings and large signals, TWIN tends to have the best minimum RMSE of all methods including the Lasso. 

Comparing the MCLLA and CD algorithms for TWIN-a and TWIN-b, we find that MCLLA tends to outperform CD in small sample size settings, however when the sample size is larger, CD performs better. This trend holds in additional simulation studies presented in the Supplementary Material.
In the Supplementary Material we present results with $p=2000$ under Models 1 and 2 and under two similar models with an increased number of active variables ($k=100$). The results are quite similar, further substantiating our theoretical results.

\section{Analysis of polymerase chain reaction (PCR) study}\label{sec:application}

\citet{lan2006combined} conducted an experiment to investigate the relationship between gene expression and gene function in mice. In the study gene expression levels were measured on  22,575 genes of 29 male and 31 female mice using Affymetrix MOE430 microarrays. To examine gene function, three phenotypes phosphoenopyruvate carboxykinase (PEPCK), glycerol-3-phosphate
acyltransferase (GPAT), and stearoyl-CoA desaturase 1 (SCD1) were measured for each of the mice by quantitative real-time PCR. The data are publicly available from the Gene Expression Omnibus (GEO) project (\url{http://www.ncbi.nlm.nih.gov/geo} via accession number GSE3330).

For ease of presentation we restrict our focus to analysis of the SCD1 phenotype, which is a key enzyme in the metabolism of fatty acids. As there is no natural validation data available for this study, we compare different methods by repeatedly drawing random splits of the 60 samples into 55 training samples and 5 testing samples. As a preprocessing step we take a log transformation of the gene expression levels. Using each comparator method we fit a model predicting SCD1 using all 22,575 gene expression levels. The sample correlations of the design matrix range from -0.83 to 0.99 with 10th and 90th quantiles of -0.22 and 0.24, respectively. Each method is evaluated by the average out-of-sample mean squared prediction error (MSPE) on the testing samples ($MSPE = S^{-1}\sum_{s = 1}^S\sum_{i \in I_{test,s}}(y_i - \bsX_i'{\widehat{\bsbeta}_{train, s}})^2/|I_{test,s}|$, where $I_{test,s}$ are the indices of the testing samples for the $s$th replication and $\widehat{\bsbeta}_{train, s}$ is an estimate of $\bsbeta$ using the training samples from the $s$th replication). We repeat this procedure $S=100$ times. 
We consider the Lasso, MCP, SCAD, and TWIN penalties in our analysis and for all methods use 10-fold cross validation for selection of the tuning parameter $\lambda$. The additional tuning parameters for all methods were chosen as described in Section \ref{sec:simulation}. Due to the small sample size, we utilize the MCLLA algorithm for TWIN. We also investigated TWIN with $\tau = 0.15$ and the results were similar. 

The average MSPE and number of selected variables for each method are reported in Table \ref{table:test_mspe}. Both TWIN-a and TWIN-b have better predictive performance than all other methods except the Lasso while retaining very parsimonious models. MCP selects about half has many genes as TWIN on average, but its MSPE is significantly worse than that of both TWIN-a and TWIN-b.  Both TWIN penalties also yield stable results across the replications. The top two genes selected by both TWIN-a and TWIN-b are the same genes and are selected in all 100 replications by both penalties. The Lasso, MCP, and SCAD all selected one of these two genes for all replications. The gene selected second most often by TWIN was selected in all replications for the Lasso, but was only selected 10 times by SCAD and was never selected by MCP.  The third most commonly selected gene for the TWIN penalties was the same gene for both TWIN-a (selected 44 times) and TWIN-b (selected 56 times). This gene was selected 88 times by the Lasso, 30 times by SCAD, and was never selected by MCP. 
%
%
\begin{table}[h]
	\center
	\resizebox{0.95\textwidth}{!}{
		\begin{tabular}{@{\extracolsep{5pt}} cccccc}
			\toprule\toprule
			Method & Lasso & MCP & SCAD & TWIN-a  & TWIN-b  \\
			\midrule
			MSPE & $0.613 (0.058)$ & $0.760 (0.048)$ & $0.740 (0.048)$ & $\mathbf{0.609} (0.040)$  & $0.651 (0.049)$ \\
			Number Selected & $40.58 (1.23)$ & $1.66 (0.10)$ & $26.16 (0.74)$ & $3.16 (0.14)$ & $3.58 (0.16)$ \\
			\bottomrule
		\end{tabular}
	}
		\caption{Average test set MSPE and number of variables selected by Lasso, MCP, SCAD, TWIN-a, and TWIN-b. Standard errors are in parentheses. Note that $n^{-1}\sum_{i=1}^n(y_i - \bar{y}) ^ 2 = 2.090$, where $\bar{y}$ is the average of the response values.}
		\label{table:test_mspe}
\end{table}

\section{Discussion}
\label{sec:discussion}

In this paper we proposed a novel class of penalties for regression problems. The desirable theoretical properties of TWIN derive from its unique shape, which acts to inflate coefficient estimates in a certain range, thus alleviating issues in selection arising from shrinkage pseudo-noise.  Probabilistic bounds for selection consistency were established under a challenging linear sparsity regime with random Gaussian designs. Minimax optimality was also established under the same data-generating regimes. 
 Empirically, TWIN shows good performance even under scenarios with strong correlations in the design, suggesting that TWIN's theoretical properties may be extendible to more realistic data-generating scenarios. Motivated by this, we expect that exploration of TWIN's theoretical behavior under designs with significant correlation may be fruitful. In this work we provided asymptotically-motivated choices for the tuning parameters, however, the development of comprehensive strategies for simultaneous selection of $\tau$ and $\lambda$ based on finite sample analysis is another interesting avenue of future research.

\bigskip
\begin{center}
{\large\bf SUPPLEMENTARY MATERIAL}
\end{center}

\begin{description}

\item[Title:] The supplementary materials contain detailed proofs for results in Section \ref{sec:selectionconsis} and Section \ref{sec:estimationproperties}. (Supplement.pdf file)

\item[R-package:] R-package TWIN containing code to perform the variable selection methods described in the article. (GNU zipped tar file)

\end{description}

\bibliographystyle{Chicago}

\bibliography{references}

\end{document}